\documentclass[final]{clv3}

\usepackage{hyperref}
\usepackage{xcolor}
\usepackage{graphicx}
\usepackage{wrapfig}
\usepackage{lscape}
\usepackage{rotating}
\usepackage{epstopdf}

\definecolor{darkblue}{rgb}{0, 0, 0.5}
\hypersetup{colorlinks=true,citecolor=darkblue, linkcolor=darkblue, urlcolor=darkblue}

\issue{1}{1}{2019}

\runningtitle{The State of NLP Literature}

\runningauthor{Saif M. Mohammad}

\begin{document}

\title{The State of NLP Literature:\\ A Diachronic Analysis of the ACL Anthology}

\author{Saif M. Mohammad\thanks{1200 Montreal Rd., Ottawa, Canada. E-mail: saif.mohammad@nrc-cnrc.gc.ca.}}
\affil{National Research Council Canada}




\maketitle

\begin{abstract}
The ACL Anthology (AA) is a digital repository of tens of thousands of articles on Natural Language Processing (NLP). This paper examines the literature as a whole to identify broad trends in productivity, focus, and impact. It presents the analyses in a series of questions and answers. The goal is to record the state of the AA literature: who and how many of us are publishing? what are we publishing on? where and in what form are we publishing? and what is the impact of our publications? The answers are usually in the form of numbers, graphs, and inter-connected visualizations. Special emphasis is laid on the demographics and inclusiveness of NLP publishing. Notably, we find that only about 30\% of first authors are female, and that this percentage has not improved since the year 2000. We also show that, on average, female first authors are cited less than male first authors, even when controlling for experience. We hope that recording citation and participation gaps across demographic groups will encourage more inclusiveness and fairness in research.
\end{abstract}

\section{Introduction}

The ACL Anthology (AA) is a digital repository of tens of thousands of articles on Natural Language Processing (NLP) / Computational Linguistics (CL).\footnote{https://www.aclweb.org/anthology/} It includes papers published in the family of ACL conferences as well as in other NLP conferences such as LREC and RANLP. AA is the largest single source of scientific literature on NLP.

This project, which we call \textit{NLP Scholar}, examines the literature as a whole to identify broad trends in productivity, focus, and impact. We will present the analyses in a sequence of questions and answers. The questions range from fairly mundane to oh-that-will-be-good-to-know. Our broader goal here is simply to record the state of the AA literature: who and how many of us are publishing? what are we publishing on? where and in what form are we publishing? and what is the impact of our publications? The answers are usually in the form of numbers, graphs, and inter-connected visualizations.

We focus on the following aspects of NLP research: size, demographics, areas of research, impact, and correlation of citations with demographic attributes (age and gender).\\

\vspace*{-1mm}
\noindent \textbf{Target Audience:} The analyses presented here are likely to be of interest to any NLP researcher. This might be particularly the case for those that are new to the field and wish to get a broad overview of the NLP publishing landscape. On the other hand, even seasoned NLP'ers 
have likely wondered about the questions raised here and might be interested in the empirical evidence.\\

\vspace*{-1mm}
 \noindent    \textbf{Data:} The analyses presented below are based on information about the papers taken directly from AA (as of June 2019) and citation information extracted from Google Scholar (as of June 2019). Thus, all subsequent papers and citations are not included in the analysis. A fresh data collection is planned for January 2020.\\

 \vspace*{-1mm}
    \noindent    \textbf{Interactive Visualizations:} The visualizations we are developing for this work (using Tableau) are interactive---so one can hover, click to select and filter, move sliders, etc. Since this work is high in the number of visualizations, the main visualizations are presented as figures in the paper and some sets of visualizations are pointed to online. The interactive visualizations and data will be made available through the first author's website after peer review.\footnote{http://saifmohammad.com}\\
 
 \vspace*{-1mm}
 \noindent \textbf{Related Work:} This work builds on past research, including that on Google Scholar \cite{khabsa2014number,howland2010scholarly,orduna2014size,martin2018google},
 on the analysis of NLP papers \cite{radev2016bibliometric,anderson2012towards,bird2008acl,schluter2018glass,mariani2018nlp4nlp,qazvinian2013generating}, 
 on citation intent \cite{aya2005citation,teufel2006automatic,pham2003new,nanba2011classification,mohammad2009using,zhu2015measuring},
 and
 on measuring scholarly impact \cite{ravenscroft2017measuring,priem2010scientometrics,bulaitis2017measuring,bos2019interdisciplinary,ioannidis2019standardized,yogatama2011predicting}.\\
    
\vspace*{-1mm}
 \noindent    \textbf{Caveats and Ethical Considerations:} 
 We list several caveats and limitations throughout the paper. A compilation of these is also available online in the About NLP Scholar page.\footnote{https://medium.com/@nlpscholar/about-nlp-scholar-62cb3b0f4488}\\
 
 \noindent The analyses presented here are also available as a series of blog posts.\footnote{https://medium.com/@nlpscholar/state-of-nlp-cbf768492f90}

\section{Size}

\textit{Q. How big is the ACL Anthology (AA)? How is it changing with time?}\\

\begin{figure*}[t]
 \begin{center}
 	\includegraphics[width=\columnwidth]{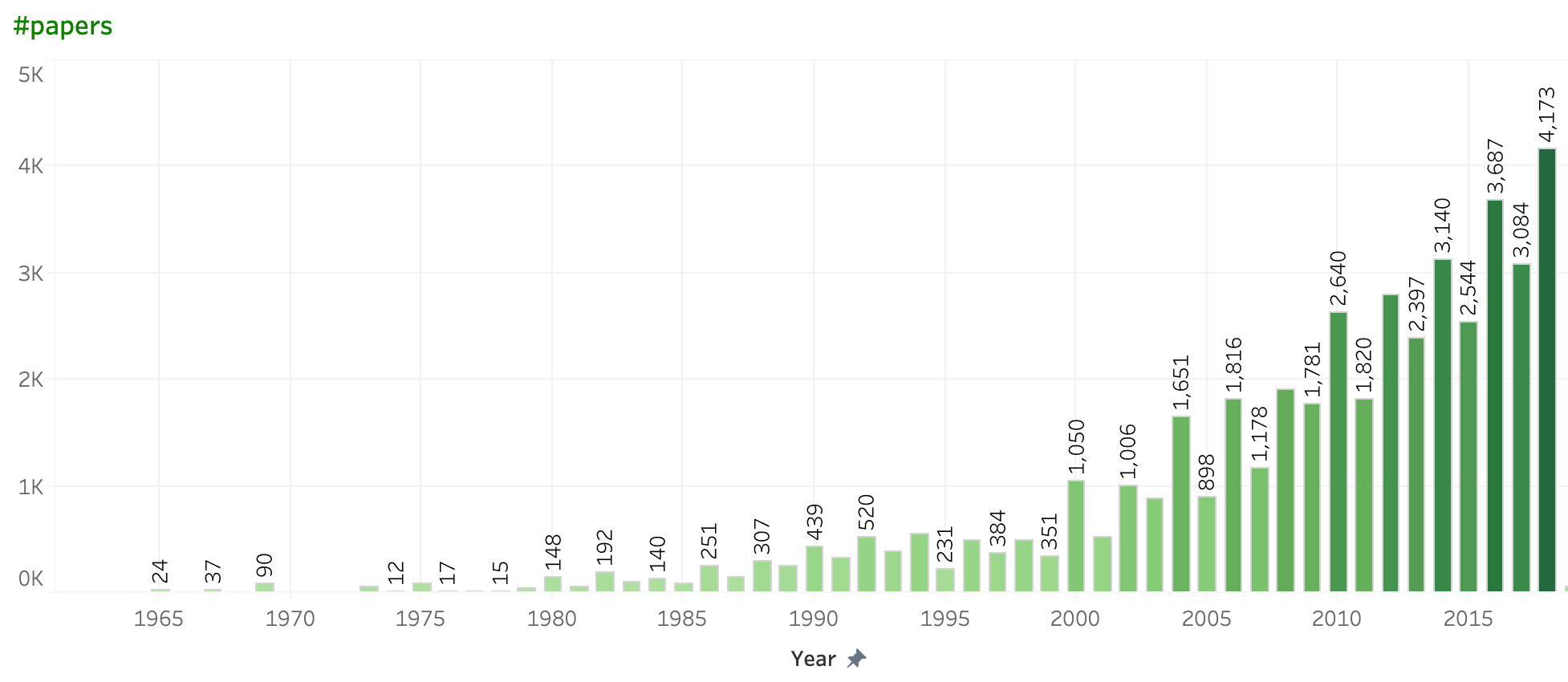}
 	\caption{The number of AA papers published in each of the years from 1965 to 2018.}
 	\label{fig:numpapersByYear}
 \end{center}
 \end{figure*}

\noindent A. As of June 2019, AA had $\sim$50K entries, however, this includes some number of entries that are not truly research publications (for example, forewords, prefaces, table of contents, programs, schedules, indexes, calls for papers/participation, lists of reviewers, lists of tutorial abstracts, invited talks, appendices, session information, obituaries, book reviews, newsletters, lists of proceedings, lifetime achievement awards, erratum, and notes). We discard them for the analyses here. (Note: CL journal includes position papers like squibs, letter to editor, opinion, etc. We do not discard them.) We are then left with 44,896 articles. Figure \ref{fig:numpapersByYear} shows a graph of the number of papers published in each of the years from 1965 to 2018.\\

\noindent Discussion: Observe that there was a spurt in the 1990s, but things really took off since the year 2000, and the growth continues. Also, note that the number of publications is considerably higher in alternate years. This is due to biennial conferences. Since 1998 the largest of such conferences has been LREC (In 2018 alone LREC had over 700 main conferences papers and additional papers from its 29 workshops). COLING, another biennial conference (also occurring in the even years) has about 45\% of the number of main conference papers as LREC.\\\\\\\\

\noindent \textit{Q. How many people publish in the ACL Anthology (NLP conferences)?}\\

\begin{figure*}[t]
 \begin{center}
 	\includegraphics[width=\columnwidth]{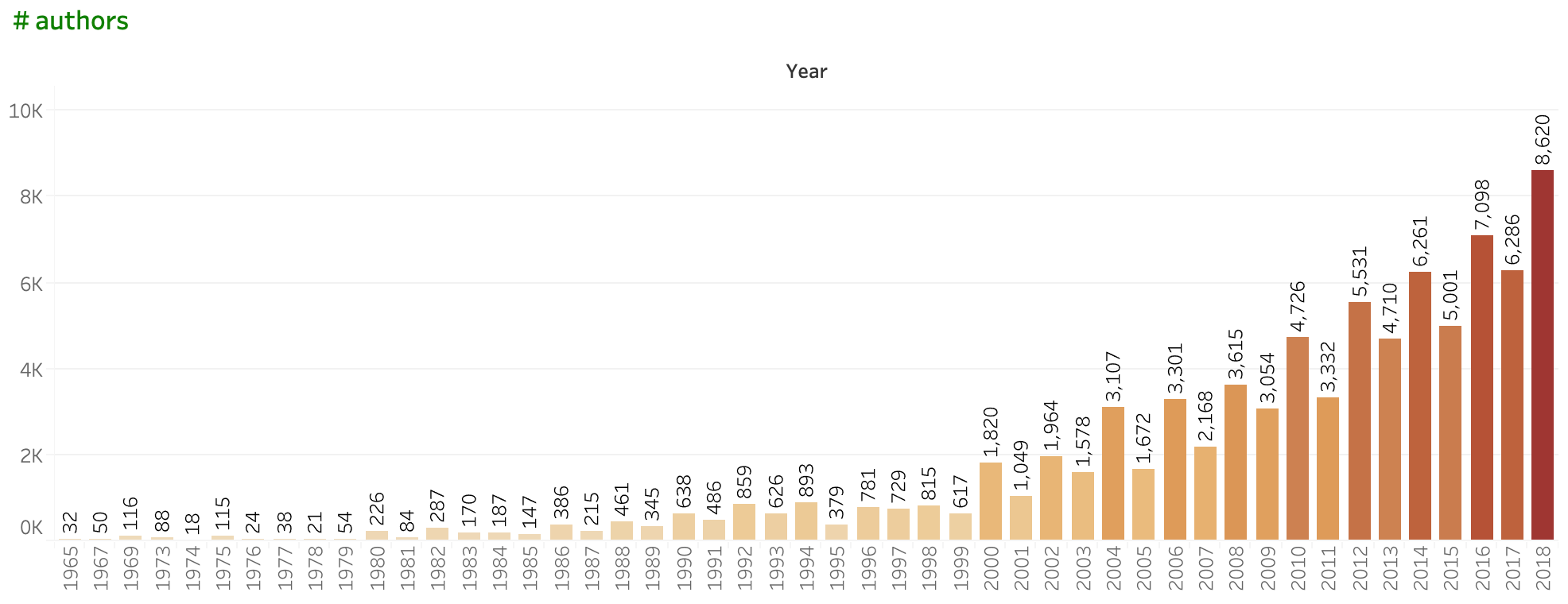}
 	\caption{The number of authors of AA papers from 1965 to 2018.}
 	\label{fig:numauthors}
 \end{center}
 \end{figure*}

\noindent A. Figure \ref{fig:numauthors} shows a graph of the number of authors (of AA papers) over the years:\\

\noindent Discussion: It is a good sign for the field to have a growing number of people join its ranks as researchers. A further interesting question would be:\\

\noindent \textit{Q. How many people are actively publishing in NLP?}\\

\noindent A. It is hard to know the exact number, but we can determine the number of people who have published in AA in the last N years.\\

\noindent \#people who published at least one paper in 2017 and 2018 (2 years): $\sim$12k (11,957 to be precise)\\
\noindent \#people who published at least one paper 2015 through 2018 (4 years):$\sim$17.5k (17,457 to be precise)\\

\noindent Of course, some number of researchers published NLP papers in non-AA venues, and some number are active NLP researchers who may not have published papers in the last few years.\\

\noindent \textit{Q. How many journal papers exist in the AA? How many main conference papers? How many workshop papers?}\\

\begin{figure*}[t]
 \begin{center}
 	\includegraphics[width=\columnwidth]{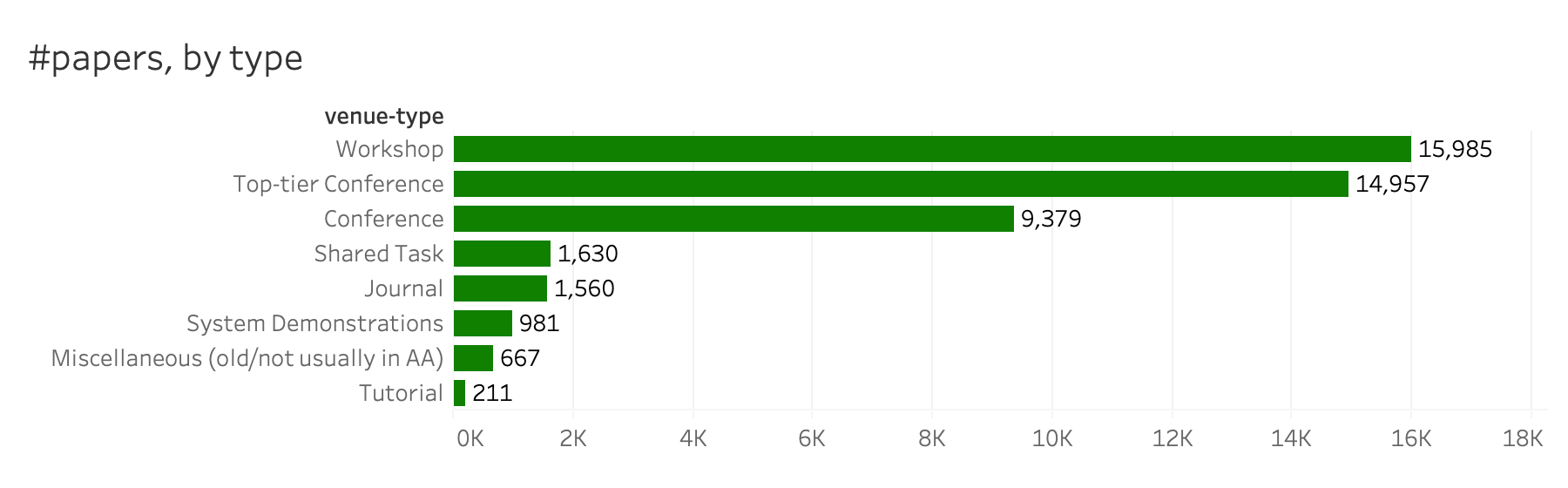}
 	\caption{Number of AA papers by type.}
 	\label{fig:NumPapersByType}
 \end{center}
 \end{figure*}

\noindent A. See Figure \ref{fig:NumPapersByType}.\\

\noindent Discussion: The number of journal papers is dwarfed by the number of conference and workshop papers. (This is common in computer science. Even though NLP is a broad interdisciplinary field, the influence of computer science practices on NLP is particularly strong.) Shared task and system demo papers are relatively new (introduced in the 2000s), but their numbers are already significant and growing.

Creating a separate class for “Top-tier Conference” is somewhat arbitrary, but it helps make certain comparisons more meaningful (for example, when comparing the average number of citations, etc.). For this work, we consider ACL, EMNLP, NAACL, COLING, and EACL as top-tier conferences, but certainly other groupings are also reasonable.\\

\noindent \textit{Q. How many papers have been published at ACL (main conference papers)? What are the other NLP venues and what is the distribution of the number of papers across various CL/NLP venues?}\\

\noindent A. \# ACL (main conference papers) as of June 2018: 4,839\\

\begin{figure*}[t]
 \begin{center}
 	\includegraphics[width=\columnwidth]{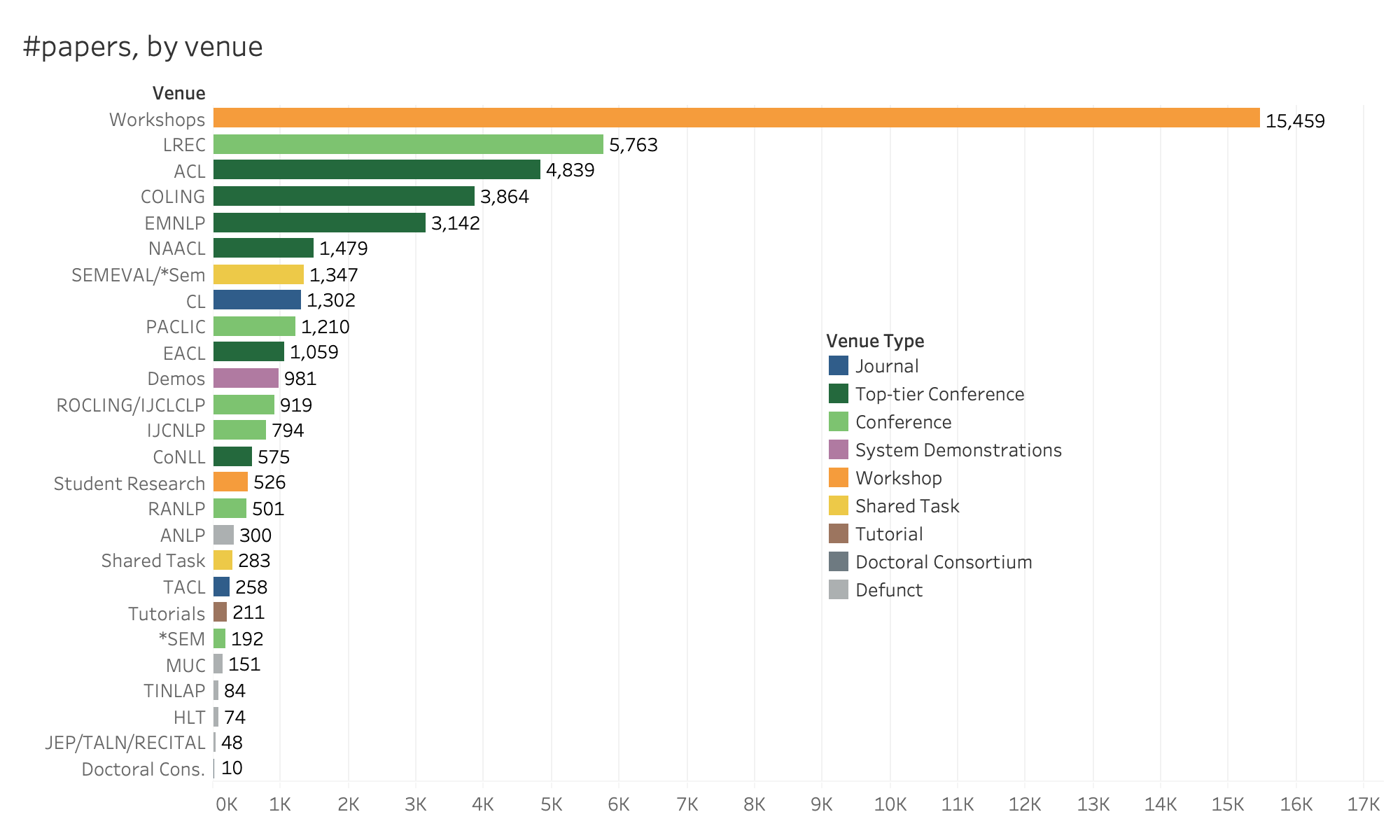}
 	\caption{The number of main conference papers for various venues and paper types (workshop papers, demos, etc.).}
 	\label{fig:NumPapersByVenue}
 \end{center}
 \end{figure*}

\noindent The same workshop can co-occur with different conferences in different years, so we grouped all workshop papers in their own class. We did the same for tutorials, system demonstration papers (demos), and student research papers. Figure \ref{fig:NumPapersByVenue} shows the number of main conference papers for various venues and paper types (workshop papers, demos, etc.).\\

\noindent Discussion: Even though LREC is a relatively new conference that occurs only once in two years, it tends to have a high acceptance rate ($\sim$60\%), and enjoys substantial participation. Thus, LREC is already the largest single source of NLP conference papers. SemEval, which started as SenseEval in 1998 and occurred once in two or three years, has now morphed into an annual two-day workshop---SemEval. It is the largest single source of NLP shared task papers.

\section{Demographics (focus of analysis: gender, age, and geographic diversity)}


NLP, like most other areas of research, suffers from poor demographic diversity. There is very little to low representation from certain nationalities, race, gender, language, income, age, physical abilities, etc. This impacts the breadth of technologies we create, how useful they are, and whether they reach those that need it most.
In this section, we analyze three specific attributes among many that deserve attention: gender (specifically, the number of women researchers in NLP), age (more precisely, the number of years of NLP paper publishing experience), and the amount of research in various languages (which loosely correlates with geographic diversity).\footnote{It should be noted that there exists very little work on tracing the participation and contributions of those with non-binary and other gender identities. Similarly, tracking the skew in authors of diverse income, experiences, and abilities is also crucial.}

\subsection{Gender}

The ACL Anthology does not record demographic information about the paper authors. (Until recently, ACL and other NLP conferences did not record demographic information of the authors.) However, many first names have strong associations with a male or female gender. We will use these names to estimate the percentage of female first authors in NLP.

The US Social Security Administration publishes a  database of names and genders of newborns.\footnote{https://www.ssa.gov/oact/babynames/limits.html} We use the dataset to identify 55,133 first names that are strongly associated with females (probability $\ge$99\%) and 29,873 first names that are strongly associated with males (probability $\ge$99\%). (As a side, it is interesting to note that there is markedly greater diversity in female names than in male names.) We identified 26,637 of the 44,896 AA papers ($\sim$60\%) where the first authors have one of these names and determine the percentage of female first author papers across the years. We will refer to this subset of AA papers as AA*.\\

\vspace*{-3mm}
\noindent Note the following caveats associated with this analysis:
\begin{enumerate}
\vspace*{-4mm}
\item    The names dataset used has a lower representation of names from nationalities other than the US. However, 
there is a large expatriate population living in the US.
\vspace*{-1mm}
\item    Chinese names (especially in the romanized form) are not good indicators of gender. Thus the method presented here disregards most Chinese names, and the results of the analysis apply to the group of researchers excluding those with Chinese names.
\vspace*{-1mm}
\item    The dataset only records names associated with two genders. 
\end{enumerate}
\vspace*{-3mm}
\noindent The approach presented here is meant to be 
an approximation in the absence of true gender information.\\


 \begin{figure*}[t]
 \begin{center}
 	\includegraphics[width=\columnwidth]{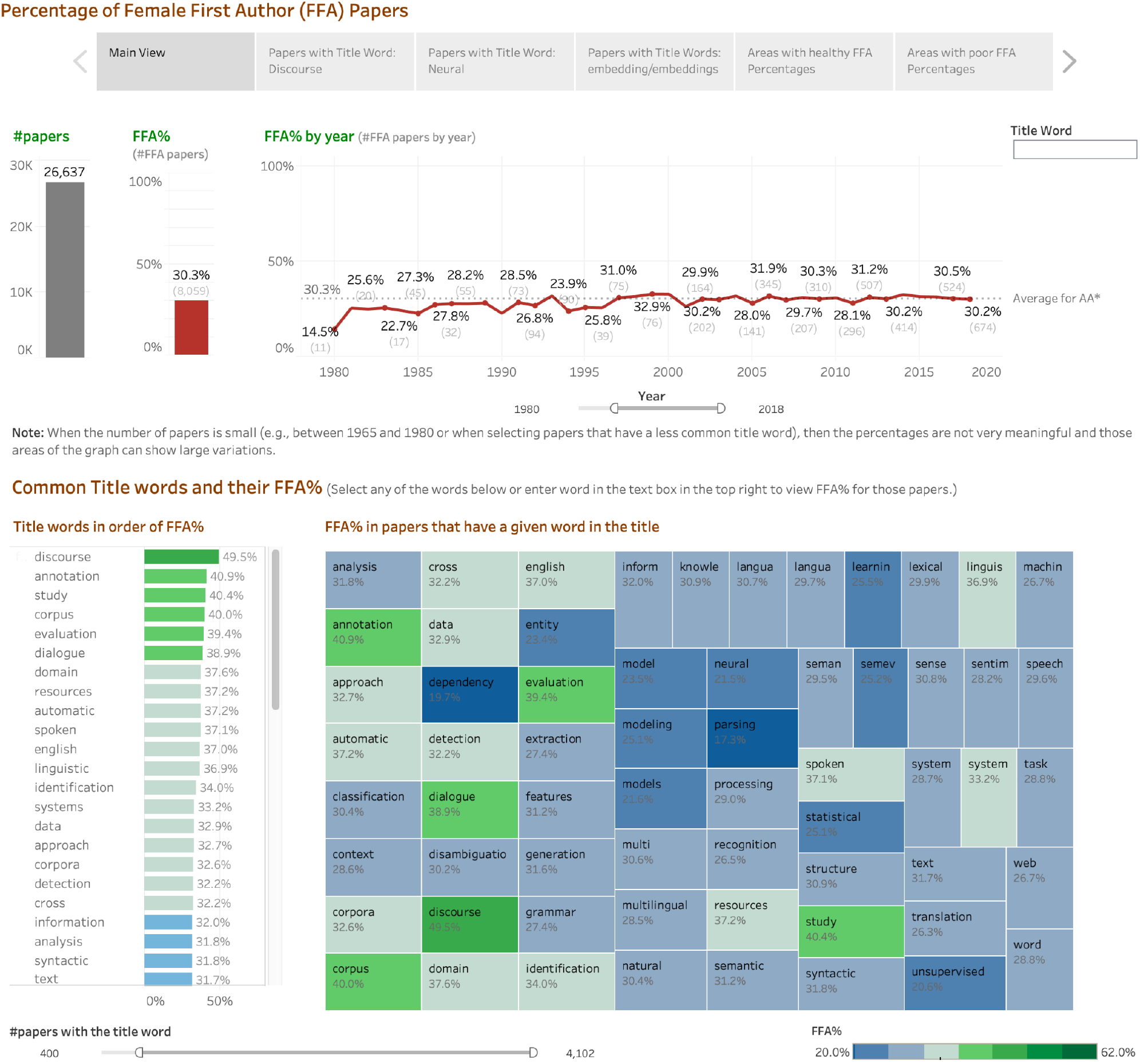}
 	\caption{Female first author (FFA) percentage over the years.}
 	\label{fig:FFAP}
 \end{center}
 \vspace*{-6mm}
 \end{figure*}

\noindent \textit{Q. What percent of the AA* papers have female first authors (FFA)? How has this percentage changed with time?}\\

\noindent A. Overall FFA\%: 30.3\%. Figure \ref{fig:FFAP} shows how FFA\% has changed with time. Common paper title words and FFA\% of papers that have those words are shown in the bottom half of the image.
Note that the slider at the bottom has been set to 400, i.e., only those title words that occur in 400 or more papers are shown. The legend on the bottom right shows that low FFA scores are shown in shades of blue, whereas relatively higher FFA scores are shown in shades of green.\\

\vspace*{-3mm}
\noindent Discussion: Observe that as a community, we are far from obtaining male-female parity in terms of first authors.
A further striking (and concerning) observation is that the female first author percentage has not improved since the years 1999 and 2000 when the FFA percentages were highest (32.9\% and 32.8\%, respectively). In fact there seems to even be a slight downward trend in recent years.
The calculations shown above are for the percentage of papers that have female first authors. The percentage of female first authors is about the same ($\sim$31\%). On average male authors had a slightly higher average number of publications than female authors.

To put these numbers in context, the percentage of female scientists world wide (considering all areas of research) has been estimated to be around 30\%. The reported percentages for many computer science sub-fields are much lower. (See Women in Science (2015).\footnote{https://unesdoc.unesco.org/ark:/48223/pf0000235155})
The percentages are much higher for certain other fields such as psychology and linguistics. 
(See this study for psychology\footnote{https://www.apa.org/gradpsych/2011/01/cover-men} and this study for linguistics\footnote{https://www.linguisticsociety.org/sites/default/files/Annual\_Report\_2016.pdf}.) If we can identify ways to move the needle on the FFA percentage and get it closer to 50\% (or more), NLP can be a beacon to many other fields, especially in the sciences.

\begin{figure*}[t!]
 \begin{center}
 	\includegraphics[width=\columnwidth]{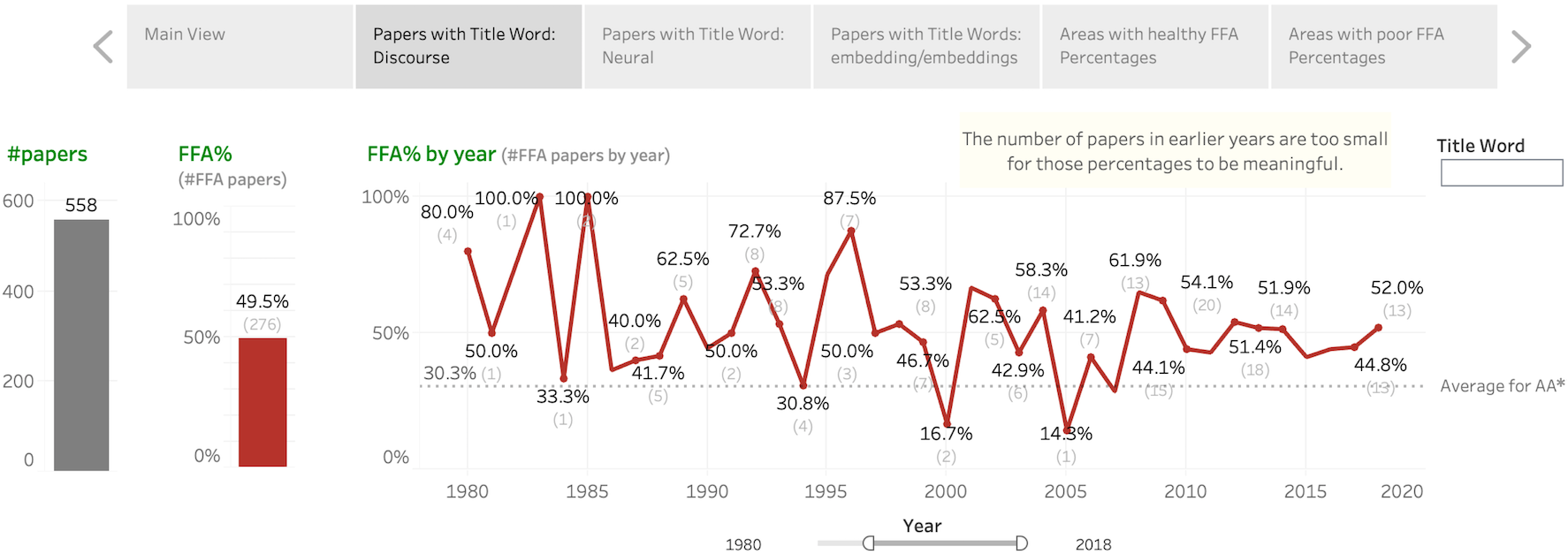}
 	\caption{FFA percentages for papers that have the word \textit{discourse} in the title.}
 	\label{fig:FFA-discourse}
 \end{center}
 \vspace*{-3mm}
 \end{figure*}

\begin{figure*}[t!]
 \begin{center}
 	\includegraphics[width=\columnwidth]{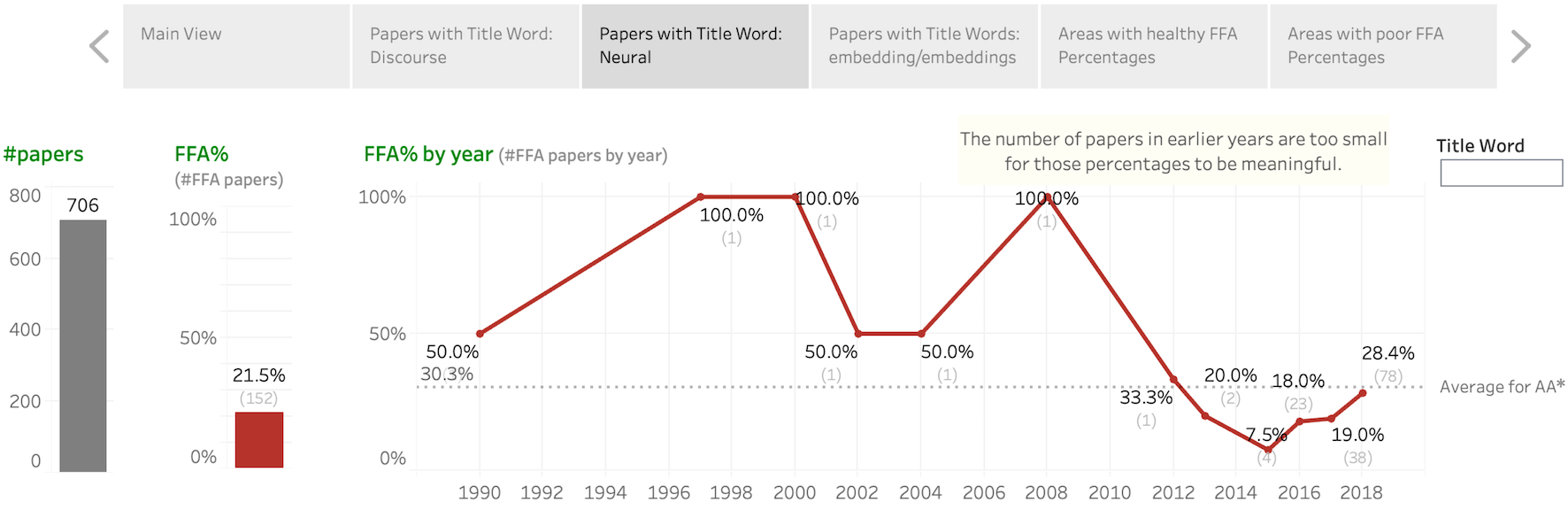}
 	\caption{FFA percentages for papers that have the word \textit{neural} in the title.}
 	\label{fig:FFA-neural}
 \end{center}
 \vspace*{-3mm}
 \end{figure*}

FFA percentages are particularly low for papers that have parsing, neural, and unsupervised in the title.
There are some areas within NLP that enjoy a healthier female-male parity in terms of first authors of papers. Figure \ref{fig:FFA-discourse} shows FFA percentages for papers that have the word \textit{discourse} in the title.
There is burgeoning research on neural NLP in the last few years. Figure \ref{fig:FFA-neural} shows FFA percentages for papers that have the word \textit{neural} in the title.

  \begin{figure*}[t!]
 \begin{center}
 	\includegraphics[width=0.5\columnwidth]{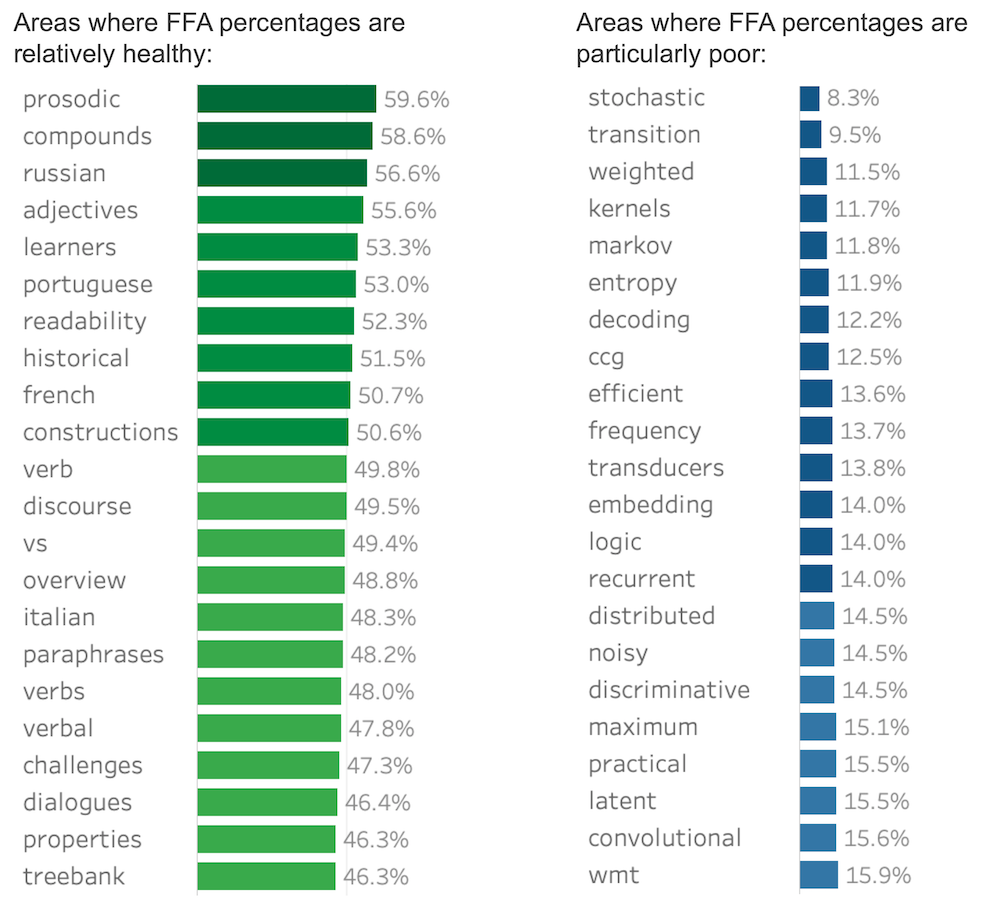}
 	\caption{Lists of terms with the highest and lowest FFA percentages, respectively.}
 	\label{fig:FFA-Good-Bad}
 \end{center}
 \vspace*{-3mm}
 \end{figure*}

 Figure \ref{fig:FFA-Good-Bad} shows lists of terms with the highest and lowest FFA percentages, respectively, when considering terms that occur in at least 50 paper titles (instead of 400 in the analysis above).
Observe that FFA percentages are relatively higher in non-English European language research such as papers on Russian, Portuguese, French, and Italian.
FFA percentages are also relatively higher for certain areas of NLP such as work on prosody, readability, discourse, dialogue, paraphrasing, and individual parts of speech such as adjectives and verbs.
FFA percentages are particularly low for papers on theoretical aspects of statistical modelling, and areas such as machine translation, parsing, and logic. The full lists of terms and FFA percentages will be made available with the rest of the data.

\subsection{Academic Age}

While the actual age of NLP researchers might be an interesting aspect to explore, we do not have that information. Thus, instead, we can explore a slightly different (and perhaps more useful) attribute: NLP academic age. We can define NLP academic age as the number of years one has been publishing in AA. So if this is the first year one has published in AA, then their NLP academic age is 1. If one published their first AA paper in 2001 and their latest AA paper in 2018, then their academic age is 18.\\

\noindent \textit{Q. How old are we? That is, what is the average NLP academic age of those who published papers in 2018? How has the average changed over the years? That is, have we been getting older or younger? What percentage of authors that published in 2018 were publishing their first AA paper?}\\

\begin{figure*}[t]
 \begin{center}
 	\includegraphics[width=\columnwidth]{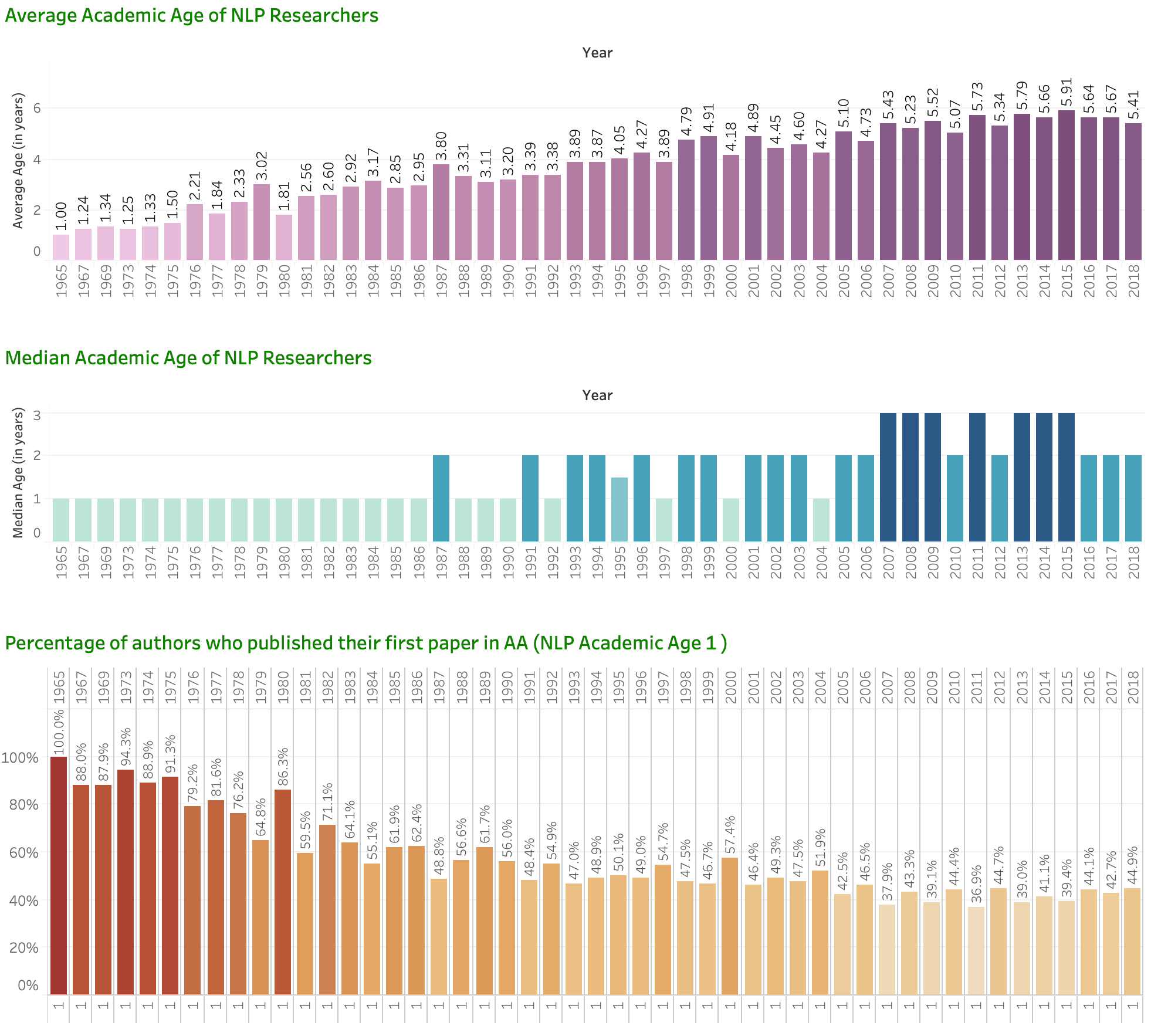}
 	\caption{Graphs showing average academic age, median academic age, and percentage of first-time publishers in AA over time.}
 	\label{fig:AAge.png}
 \end{center}
  \vspace*{-6mm}
 \end{figure*}

\noindent A. Average NLP Academic Age of people that published in 2018: 5.41 years\\
Median NLP Academic Age of people that published in 2018: 2 years\\
Percentage of 2018 authors that published their first AA paper in 2018: 44.9\%\\
Figure \ref{fig:AAge.png} shows how these numbers have changed over the years.\\

\noindent Discussion: Observe that the Average academic age has been steadily increasing over the years until 2016 and 2017, when the trend has shifted and the average academic age has started to decrease. The median age was 1 year for most of the 1965 to 1990 period, 2 years for most of the 1991 to 2006 period, 3 years for most of the 2007 to 2015 period, and back to 2 years since then. The first-time AA author percentage decreased until about 1988, after which it sort of stayed steady at around 48\% until 2004 with occasional bursts to $\sim$56\%. Since 2005, the first-time author percentage has gone up and down every other year. It seems that the even years (which are also LREC years) have a higher first-time author percentage. Perhaps, this oscillation in first-time authors percentage is related to LREC’s high acceptance rate.\\

 \begin{figure*}[t!]
 \begin{center}
 	\includegraphics[width=0.6\columnwidth]{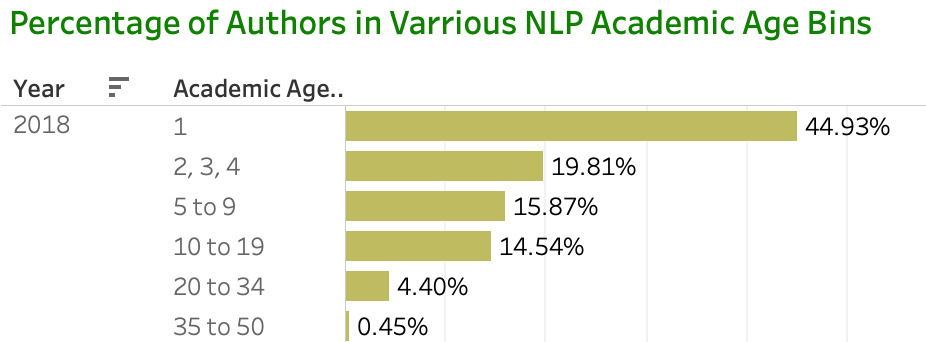}
 	\caption{The distribution of authors in academic age bins for papers published 2011--2018.}
 	\label{fig:AgeBins}
 \end{center}
  \vspace*{-9mm}
 \end{figure*}

\noindent \textit{Q. What is the distribution of authors in various academic age bins? For example, what percentage of authors that published in 2018 had an academic age of 2, 3, or 4? What percentage had an age between 5 and 9? And so on?}\\

\noindent A. See Figure \ref{fig:AgeBins}.\\

\noindent Discussion: Observe that about 65\% of the authors that published in 2018 had an academic age of less than 5.
This number has steadily reduced since 1965, was in the 60 to 70\% range in 1990s, rose to the 70 to 72\% range in early 2000s, then declined again until it reached the lowest value ($\sim$60\%) in 2010, and has again steadily risen until 2018 (65\%). Thus, even though it may sometimes seem at recent conferences that there is a large influx of new people into NLP (and that is true), proportionally speaking, the average NLP academic age is higher (more experienced) than what it has been in much of its history.

\subsection{Location (Languages)}


Automatic systems with natural language abilities are growing to be increasingly pervasive in our lives. Not only are they sources of mere convenience, but are crucial in making sure large sections of society and the world are not left behind by the information divide.
Thus, the limits of what automatic systems can do in a language, limit the world for the speakers of that language.

We know that much of the research in NLP is on English or uses English datasets. Many reasons have been proffered, and we will not go into that here. Instead, we will focus on estimating how much research pertains to non-English languages.

We will make use of the idea that often when work is done focusing on a non-English language, then the language is mentioned in the title. We collected a list of 122 languages indexed by Wiktionary and looked for the presence of these words in the titles of AA papers. (Of course there are hundreds of other lesser known languages as well, but here we wanted to see the representation of these more prominent languages in NLP literature.)

\begin{figure*}[t]
 \begin{center}
 	\includegraphics[width=\columnwidth]{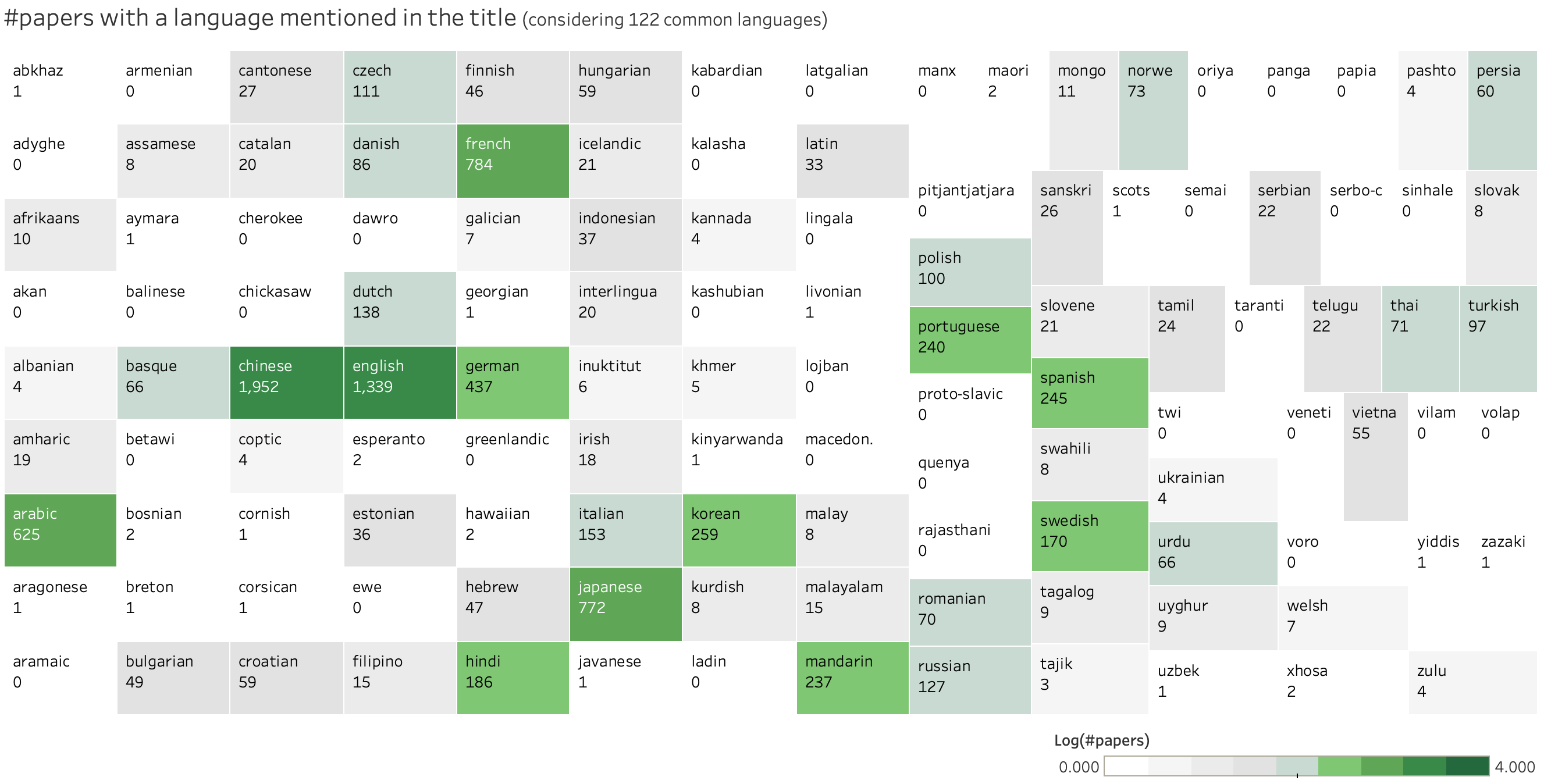}
 	\caption{A treemap of the 122 languages arranged alphabetically and shaded such that languages that appear more often in AA paper titles have a darker shade of green.}
 	\label{fig:languages}
 \end{center}
 \vspace*{-6mm}
 \end{figure*}

Figure \ref{fig:languages} is a treemap of the 122 languages arranged alphabetically and shaded such that languages that appear more often in AA paper titles have a darker shade of green.\\

\noindent Discussion: Even though the amount of work done on English is much larger than that on any other language, often the word English does not appear in the title, and this explains why English is not the first (but the second-most) common language name to appear in the titles. This is likely due to the fact that many papers fail to mention the language of study or the language of the datasets used if it is English. There is growing realization in the community that this is not quite right. However, the language of study can be named in other less prominent places than the title, for example the abstract, introduction, or when the datasets are introduced, depending on how central it is to the paper.

We can see from the treemap that the most widely spoken Asian and Western European languages enjoy good representation in AA. These include: Chinese, Arabic, Korean, Japanese, and Hindi (Asian) as well as French, German, Swedish, Spanish, Portuguese, and Italian (European). This is followed by the relatively less widely spoken European languages (such as Russian, Polish, Norwegian, Romanian, Dutch, and Czech) and Asian languages (such as Turkish, Thai, and Urdu). Most of the well-represented languages are from the Indo-European language family.
Yet, even in the limited landscape of the most common 122 languages, vast swathes are barren with inattention. Notable among these is the extremely low representation of languages from Africa, languages from non-Indo-European language families, and Indigenous languages from around the world.

\section{Areas of Research}

Natural Language Processing addresses a wide range of research questions and tasks pertaining to language and computing. It encompasses many areas of research that have seen an ebb and flow of interest over the years. In this section, we examine the terms that have been used in the titles of ACL Anthology (AA) papers. The terms in a title are particularly informative because they are used to clearly and precisely convey what the paper is about. Some journals ask authors to separately include keywords in the paper or in the meta-information, but AA papers are largely devoid of this information. Thus titles are an especially useful source of keywords for papers---keywords that are often indicative of the area of research. 

Keywords could also be extracted from abstracts and papers; we leave that for future work.
Further work is also planned on inferring areas of research using word embeddings, techniques from topic modelling, and clustering. There are clear benefits to performing analyses using that information. However, those approaches can be sensitive to the parameters used. Here, we keep things simple and explore counts of terms in paper titles. Thus the results are easily reproducible and verifiable.

\textit{Caveat:} Even though there is an association between title terms and areas of research, the association can be less strong for some terms and areas. We use the association as one (imperfect) source of information about areas of research. This information may be combined with other sources of information to draw more robust conclusions.

\noindent \textbf{Title Terms:} The title has a privileged position in a paper. It serves many functions, and here are three key ones (from an article by Sneha Kulkarni):
"A good research paper title:
1. Condenses the paper's content in a few words
2. Captures the readers' attention
3. Differentiates the paper from other papers of the same subject area".

If we examine the titles of papers in the ACL Anthology, we would expect that because of Function 1 many of the most common terms will be associated with the dominant areas of research. Function 2 (or attempting to have a catchy title) on the other hand, arguably leads to more unique and less frequent title terms. Function 3 seems crucial to the effectiveness of a title; and while at first glance it may seem like this will lead to unique title terms, often one needs to establish a connection with something familiar in order to convey how the work being presented is new or different.

It is also worth noting that a catchy term today, will likely not be catchy tomorrow. Similarly, a distinctive term today, may not be distinctive tomorrow. For example, early papers used neural in the title to distinguish themselves from non-nerual approaches, but these days neural is not particularly discriminative as far as NLP papers go.

Thus, competing and complex interactions are involved in the making of titles. Nonetheless, an arguable hypothesis is that:
broad trends in interest towards an area of research will be reflected, to some degree, in the frequencies of title terms associated with that area over time.
However, even if one does not believe in that hypothesis, it is worth examining the terms in the titles of tens of thousands of papers in the ACL Anthology---spread across many decades.\\

\noindent \textit{Q. What terms are used most commonly in the titles of the AA papers? How has that changed with time?}\\

\begin{figure*}[t]
 \begin{center}
 	\includegraphics[width=\columnwidth]{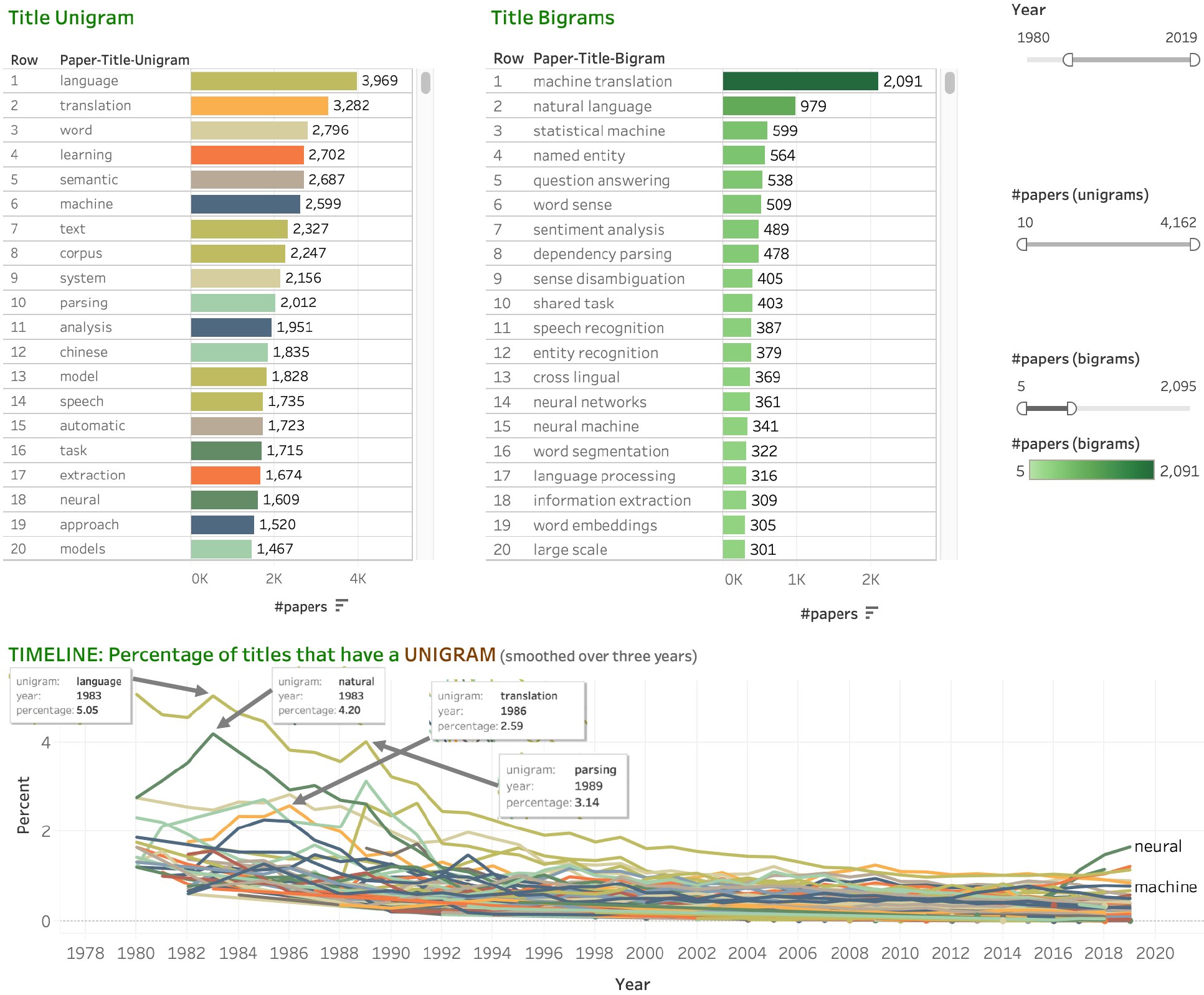}
 	\caption{The most common unigrams  and bigrams  in the titles of AA papers published 1980--2019.}
 	\label{fig:uni-bi-labels}
 \end{center}
 \vspace*{-6mm}
 \end{figure*}

\noindent A. Figure \ref{fig:uni-bi-labels} shows the most common unigrams (single word) and bigrams (two-word sequences) in the titles of papers published from 1980 to 2019. (Ignoring function words.) The timeline graph at the bottom shows the percentage of occurrences of the unigrams over the years (the colors of the unigrams in the Timeline match those in the Title Unigram list).
Note: For a given year, the timeline graph includes a point for a unigram if the sum of the frequency of the unigram in that year and the two years before it is at least ten. The period before 1980 is not included because of the small number of papers.\\

\noindent Discussion: Appropriately enough, the most common term in the titles of NLP papers is \textit{language}. Presence of high-ranking terms pertaining to machine translation suggest that it is the area of research that has received considerable attention.
Other areas associated with the high-frequency title terms include lexical semantics, named entity recognition, question answering, word sense disambiguation, and sentiment analysis. In fact, the common bigrams in the titles often correspond to names of NLP research areas. Some of the bigrams like shared task and large scale are not areas of research, but rather mechanisms or trends of research that apply broadly to many areas of research. The unigrams, also provide additional insights, such as the interest of the community in Chinese language, and in areas such as speech and parsing.

The Timeline graph is crowded in this view, but clicking on a term from the unigram list will filter out all other lines from the timeline. This is especially useful for determining whether the popularity of a term is growing or declining. (One can already see from above that neural has broken away from the pack in recent years.) Since there are many lines in the Timeline graph, Tableau labels only some (you can see neural and machine). However, hovering over a line, in the eventual interactive visualization, will display the corresponding term---as shown in the figure.

Despite being busy, the graph sheds light on the relative dominance of the most frequent terms and how that has changed with time. The vocabulary of title words is smaller when considering papers from the 1980's than in recent years. (As would be expected since the number of papers then was also relatively fewer.) Further, dominant terms such as language and translation accounted for a higher percentage than in recent years where there is a much larger diversity of topics and the dominant research areas are not as dominant as they once were.\\

  \begin{figure*}[t!]
 \begin{center}
 	\includegraphics[width=0.85\columnwidth]{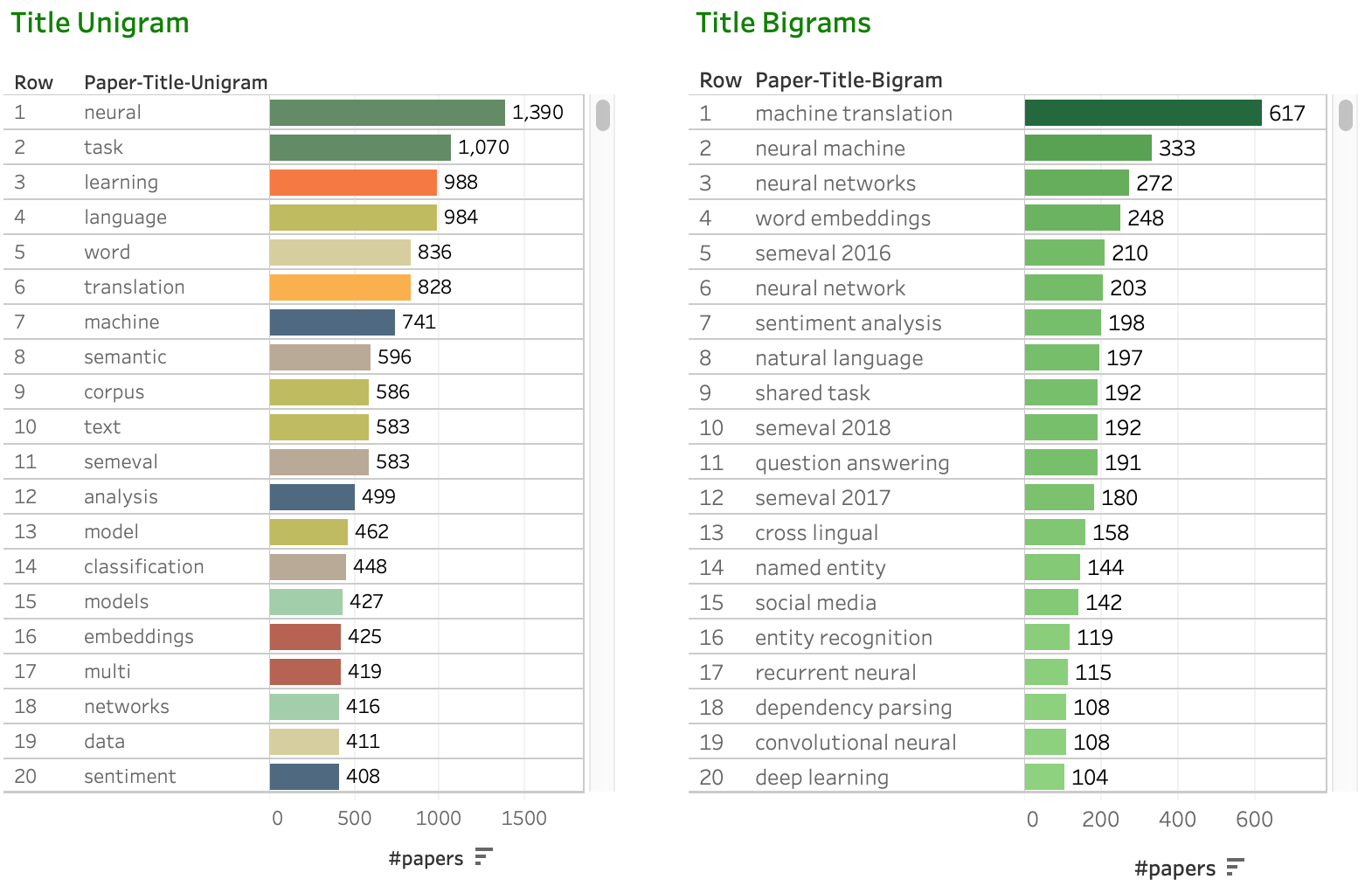}
 	\caption{The most frequent unigrams and bigrams in the titles of papers published 2016 Jan to 2019 June (time of data collection).}
 	\label{fig:2016-19-Uni-Bi-Lists}
 \end{center}
 \vspace*{-3mm}
 \end{figure*}

\noindent \textit{Q. What are the most frequent unigrams and bigrams in the titles of recent papers?}\\

\noindent A. Figure \ref{fig:2016-19-Uni-Bi-Lists} shows the most frequent unigrams and bigrams in the titles of papers published 2016 Jan to 2019 June (time of data collection).\\

\noindent Discussion: Some of the terms that have made notable gains in the top 20 unigrams and bigrams lists in recent years include: neural machine (presumably largely due to the phrase neural machine translation), neural network(s), word embeddings, recurrent neural, deep learning and the corresponding unigrams (neural, networks, etc.). We also see gains for terms related to shared tasks such as SemEval and task.\\

\noindent The sets of most frequent unigrams and bigrams in the titles of AA papers from various time spans are available
online.\footnote{https://medium.com/@nlpscholar/the-state-of-nlp-literature-part-ii-aece7bf5bad6}
Apart from clicking on terms, one can also enter the query (say parsing) in the search box at the bottom. Apart from filtering the timeline graph (bottom), this action also filters the unigram list (top left) to provide information only about the search term. This is useful because the query term may not be one of the visible top unigrams.\\ 

\begin{figure*}[t!]
 \begin{center}
 	\includegraphics[width=\columnwidth]{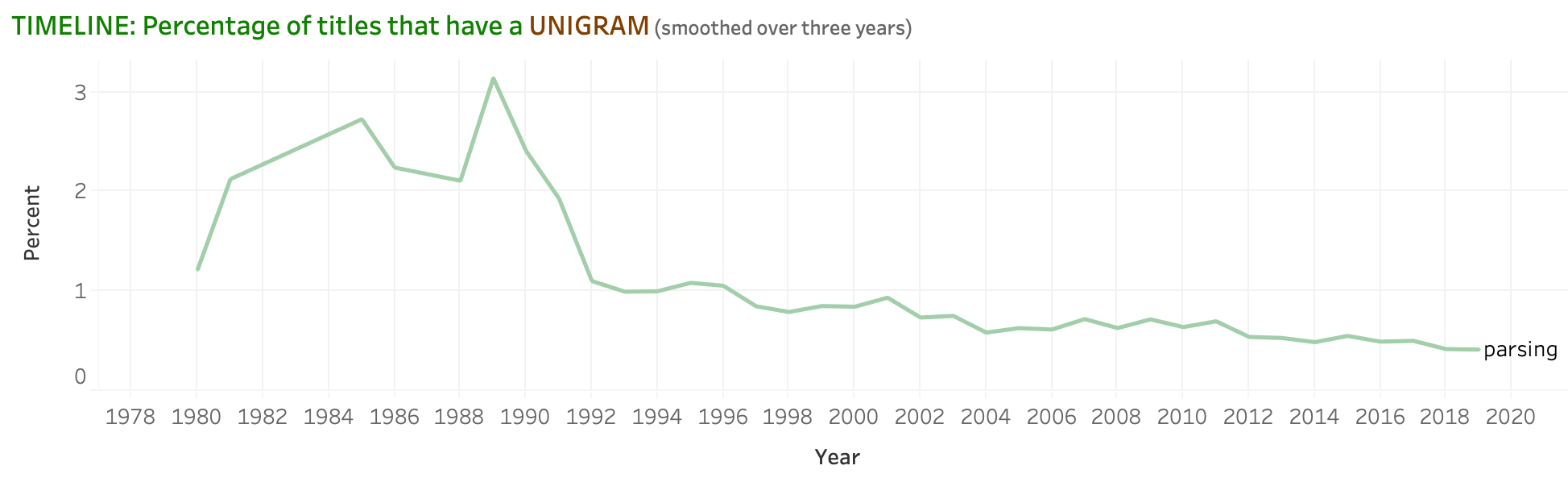}
 	\caption{The timeline graph for \textit{parsing}.}
 	\label{fig:uni-parsing}
 \end{center}
 \vspace*{-3mm}
 \end{figure*}

\noindent Figure\ref{fig:uni-parsing} shows the timeline graph for \textit{parsing}.\\

  \begin{figure*}[t!]
 \begin{center}
  	\includegraphics[width=\columnwidth]{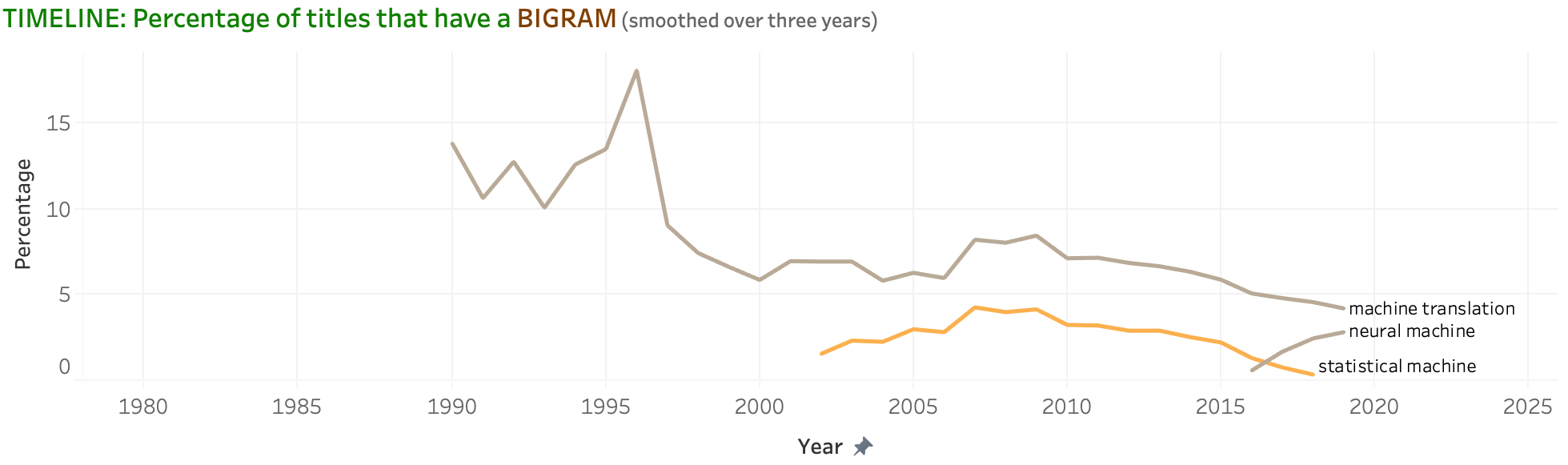}
 	\caption{The timelines for three bigrams \textit{statistical machine, neural machine,} and \textit{machine translation}.}
 	\label{fig:bi-mt}
 \end{center}
 \vspace*{-3mm}
 \end{figure*}

\noindent Discussion: Parsing seems to have enjoyed considerable attention in the 1980s, began a period of steep decline in the early 1990s, and a period of gradual decline ever since. 
One can enter multiple terms in the search box or shift/command click multiple terms to show graphs for more than one term.\\


\noindent Figure\ref{fig:bi-mt} shows the timelines for three bigrams \textit{statistical machine, neural machine,} and \textit{machine translation}:\\

\noindent Discussion: The graph indicates that there was a spike in machine translation papers in 1996, but the number of papers dropped substantially after that. Yet, its numbers have been comparatively much higher than other terms. One can also see the rise of statistical machine translation in the early 2000s followed by its decline with the rise of neural machine translation.








\section{Impact}


Research articles can have impact in a number of ways---pushing the state of the art, answering crucial questions, finding practical solutions that directly help people, making a new generation of potential-scientists excited about a field of study, and more. As scientists, it seems attractive to quantitatively measure scientific impact, and this is particularly appealing to governments and funding agencies; however, it should be noted that individual measures of research impact are limited in scope---they measure only some kinds of contributions.
Citations

The most commonly used metrics of research impact are derived from citations. A citation of a scholarly article is the explicit reference to that article. Citations serve many functions. However, a simplifying assumption is that regardless of the reason for citation, every citation counts as credit to the influence or impact of the cited work. Thus several citation-based metrics have emerged over the years including: number of citations, average citations, h-index, relative citation ratio, and impact factor.

It is not always clear why some papers get lots of citations and others do not.
  One can argue that highly cited papers have captured the imagination of the field: perhaps because they were particularly creative, opened up a new area of research, pushed the state of the art by a substantial degree, tested compelling hypotheses, or produced useful datasets, among other things.

Note however, that the number of citations is not always a reflection of the quality or importance of a piece of work. Note also that there are systematic biases that prevent certain kinds of papers from accruing citations, especially when the contributions of a piece of work are atypical, not easily quantified, or in an area where the number of scientific publications is low. Further, the citations process can be abused, for example, by egregious self-citations.

Nonetheless, given the immense volume of scientific literature, the relative ease with which one can track citations using services such as Google Scholar and Semantic Scholar, and given the lack of other easily applicable and effective metrics, citation analysis is an imperfect but useful window into research impact.

In this section, we examine citations of AA papers. We focus on two aspects:
\begin{itemize}
    \item \textit{Most cited papers:} 
    We begin by looking at the most cited papers overall and in various time spans. We will then look at most cited papers by paper-type (long, short, demo, etc) and venue (ACL, LREC, etc.). Perhaps these make interesting reading lists. Perhaps they also lead to a qualitative understanding of the kinds of AA papers that have received lots of citations.
    \item \textit{Aggregate citation metrics by time span, paper type, and venue:} Access to citation information allows us to calculate aggregate citation metrics such as average and median citations of papers published in different time periods, published in different venues, etc. These can help answer questions such as: on average, how well cited are papers published in the 1990s? on average, how many citations does a short paper get? how many citations does a long paper get? how many citations for a workshop paper? etc.
\end{itemize}


\noindent \textbf{Data:} The analyses presented below are based on information about the papers taken directly from AA (as of June 2019) and citation information extracted from Google Scholar (as of June 2019). We extracted citation information from Google Scholar profiles of authors who had a Google Scholar Profile page and had published at least three papers in the ACL Anthology. This yielded citation information for about 75\% of the papers (33,051 out of the 44,896 papers). We will refer to this subset of the ACL Anthology papers as \textit{AA’}. All citation analysis below is on AA’.



\subsection{\#Citations and Most Cited Papers}

\noindent \textit{Q. How many citations have the AA’ papers received? How is that distributed among the papers published in various decades?}\\

\begin{figure*}[t!]
 \begin{center}
 	\includegraphics[width=\columnwidth]{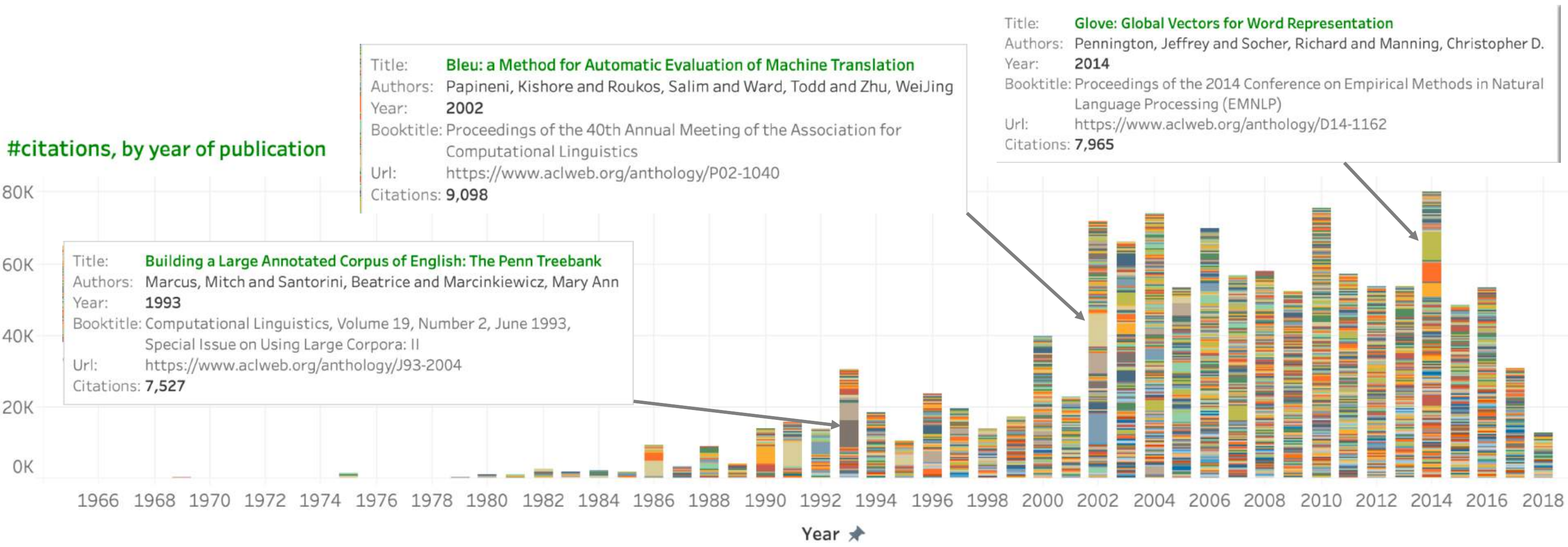}
 	\caption{A timeline graph where each year has a bar with height corresponding to the number of citations received by papers published in that year. The bar has colored fragments corresponding to each of the papers and the height of a fragment (paper) is proportional to the number of citations it has received.}
 	\label{fig:pubs-timeline-labels}
 \end{center}
 \end{figure*}

\noindent A. $\sim$1.2 million citations (as of June 2019). Figure \ref{fig:pubs-timeline-labels} shows a timeline graph where each year has a bar with height corresponding to the number of citations received by papers published in that year. Further, the bar has colored fragments corresponding to each of the papers and the height of a fragment (paper) is proportional to the number of citations it has received. Thus it is easy to spot the papers that received a large number of citations, and the years when the published papers received a large number of citations. Hovering over individual papers reveals an information box showing the paper title, authors, year of publication, publication venue, and \#citations.\\

\noindent Discussion: With time, not only have the number of papers grown, but also the number of high-citation papers. We see a marked jump in the 1990s over the previous decades, but the 2000s are the most notable in terms of the high number of citations. The 2010s papers will likely surpass the 2000s papers in the years to come.\\

\noindent \textit{Q. What are the most cited papers in AA'?}\\

\begin{figure*}[t!]
 \begin{center}
 	\includegraphics[width=\columnwidth]{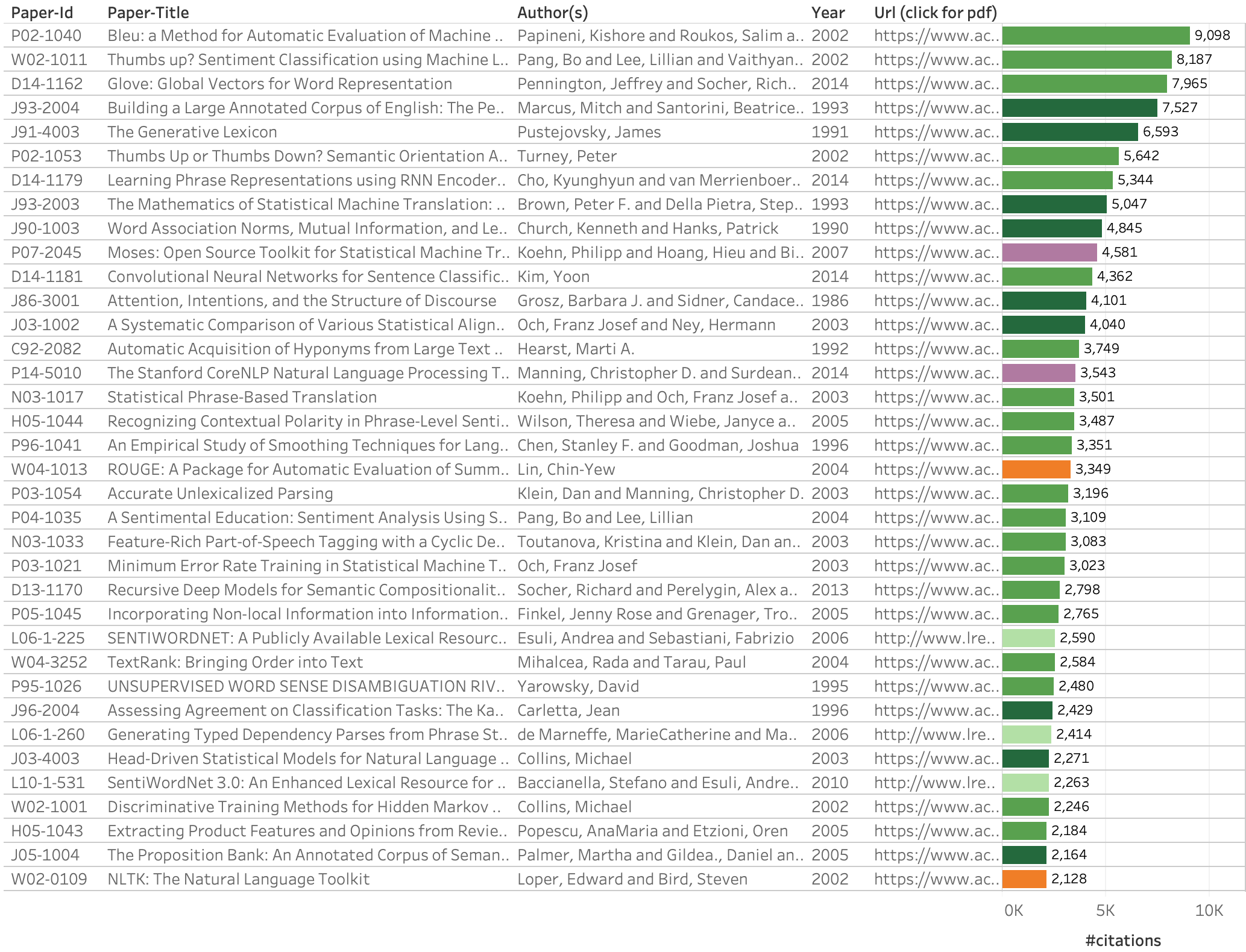}
 	\caption{The most cited papers in AA'.}
 	\label{fig:MC-All}
 \end{center}
 \vspace*{-6mm}
 \end{figure*}

\noindent  A. Figure \ref{fig:MC-All} shoes the most cited papers in the AA'.

\noindent Discussion: We see that the top-tier conference papers (green) are some of the most cited papers in AA’. There are a notable number of journal papers (dark green) in the most cited list as well, but very few demo (purple) and workshop (orange) papers.

    In the interactive visualizations (to be released later), one can click on the url to be to taken directly to the paper’s landing page in the ACL Anthology website. That page includes links to meta information, the pdf, and associated files such as videos and appendices. There will also be functionality to download the lists. Alas, copying the lists from the screenshots shown here is not easy.\\

\noindent \textit{Q. What are the most cited AA' journal papers ? What are the most cited AA' workshop papers? What are the most cited AA' shared task papers? What are the most cited AA' demo papers? What are the most cited tutorials?}\\

\noindent A. The most cited AA’ journal papers, conference papers, workshop papers, system demo papers, shared task papers, and tutorials can be viewed online.\footnote{https://medium.com/@nlpscholar/the-state-of-nlp-literature-part-iiia-845eb5dc3364}
The most cited papers from individual venues (ACL, CL journal, TACL, EMNLP, LREC, etc.) can also be viewed there.\\

\noindent Discussion: Machine translation papers are well-represented in many of these lists, but especially in the system demo papers list. Toolkits such as MT evaluation ones, NLTK, Stanford Core NLP, WordNet Similarity, and OpenNMT have highly cited demo or workshop papers.

The shared task papers list is dominated by task description papers (papers by task organizers describing the data and task), especially for sentiment analysis tasks. However, the list also includes papers by top-performing systems in these shared tasks, such as the NRC-Canada, HidelTime, and UKP papers.\\


\noindent \textit{Q. What are the most cited AA' papers in the last decade?}\\

\begin{figure*}[t!]
 \begin{center}
 	\includegraphics[width=\columnwidth]{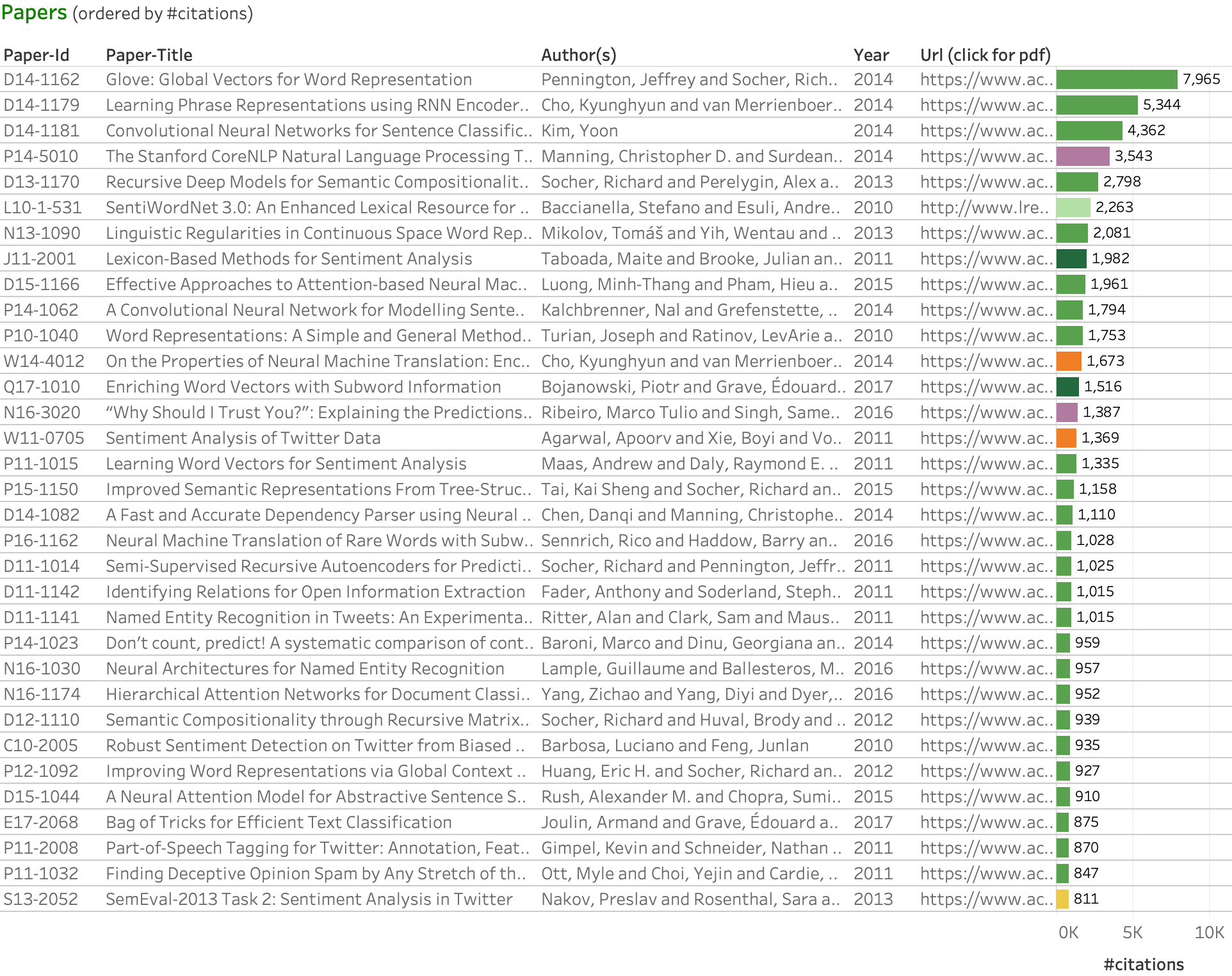}
 	\caption{The most cited AA' papers in the 2010s.}
 	\label{fig:MC-10-19}
 \end{center}
  \vspace*{-9mm}
 \end{figure*}

\noindent A. Figure \ref{fig:MC-10-19} shows the most cited AA' papers in the 2010s. The most cited AA' papers from the earlier periods are available online.\\

\noindent Discussion: The early period (1965–1989) list includes papers focused on grammar and linguistic structure. The 1990s list has papers addressing many different NLP problems with statistical approaches. Papers on MT and sentiment analysis are frequent in the 2000s list. The 2010s are dominated by papers on word embeddings and neural representations.

\subsection{Average Citations by Time Span}

\noindent \textit{Q. How many citations did the papers published between 1990 and 1994 receive? What is the average number of citations that a paper published between 1990 and 1994 has received? What are the numbers for other time spans?}\\

\begin{figure*}[t!]
 \begin{center}
 	\includegraphics[width=\columnwidth]{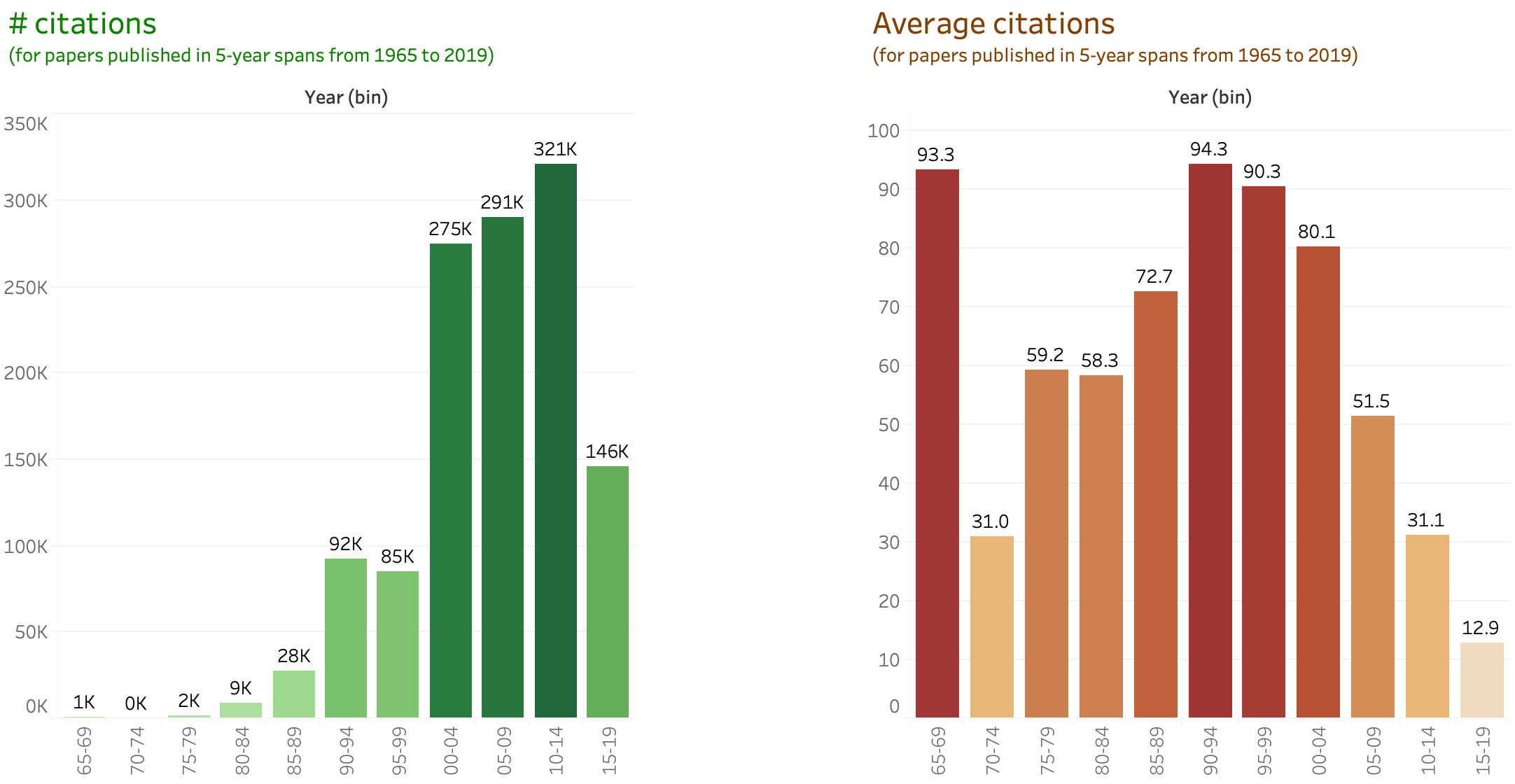}
 	\caption{Left-side graph: Total number of citations received by AA’ papers in various 5-year time spans. Right-side graph 2: Average citations per paper from various time spans.}
 	\label{fig:citationsByLustrum}
 \end{center}
 \end{figure*}

\noindent A. Total citations for papers published between 1990 and 1994: $\sim$92k\\
Average citations for papers published between 1990 and 1994: 94.3\\
Figure \ref{fig:citationsByLustrum} shows the numbers for various time spans.\\

\noindent Discussion: The early 1990s were an interesting period for NLP with the use of data from the World Wide Web and technologies from speech processing. This was the period with the highest average citations per paper, closely followed by the 1965–1969 and 1995–1999 periods. The 2000–2004 period is notable for:
(1) a markedly larger number of citations than the previous decades;
(2) third highest average number of citations. The drop off in the average citations for recent 5-year spans is largely because they have not had as much time to collect citations.

\begin{figure*}[t!]
 \begin{center}
 	\includegraphics[width=0.9\columnwidth]{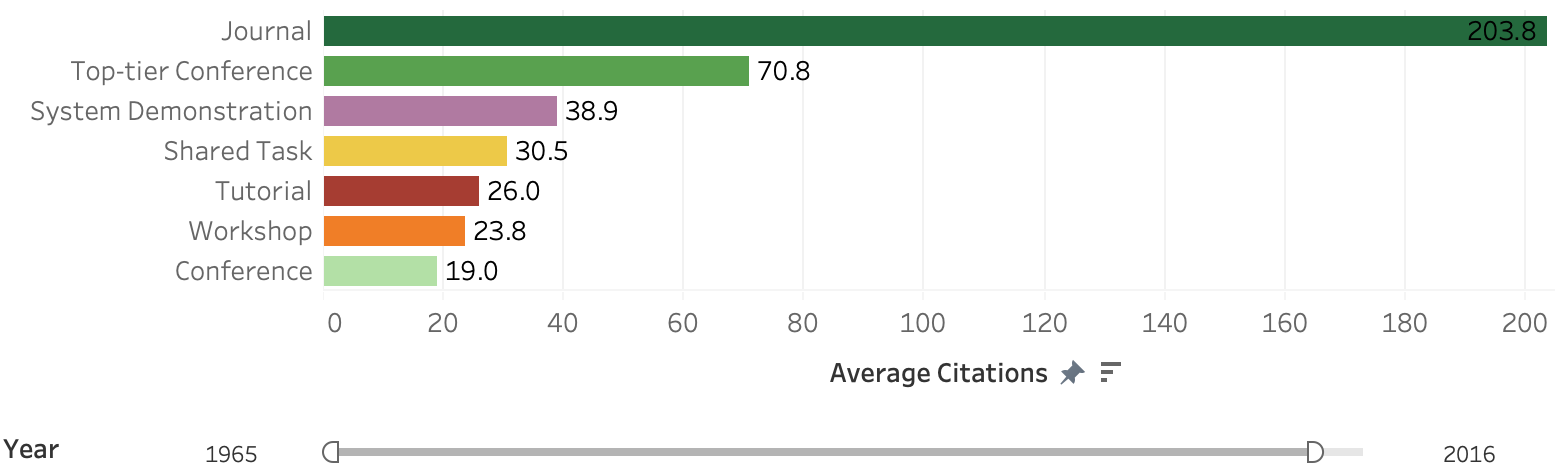}
 	\caption{Average citations by paper type when considering papers published 1965–2016.}
 	\label{fig:PType-AvgCitn-1965-2016}
 \end{center}
 \end{figure*}
 

\begin{figure*}[t!]
 \begin{center}
 	\includegraphics[width=0.9\columnwidth]{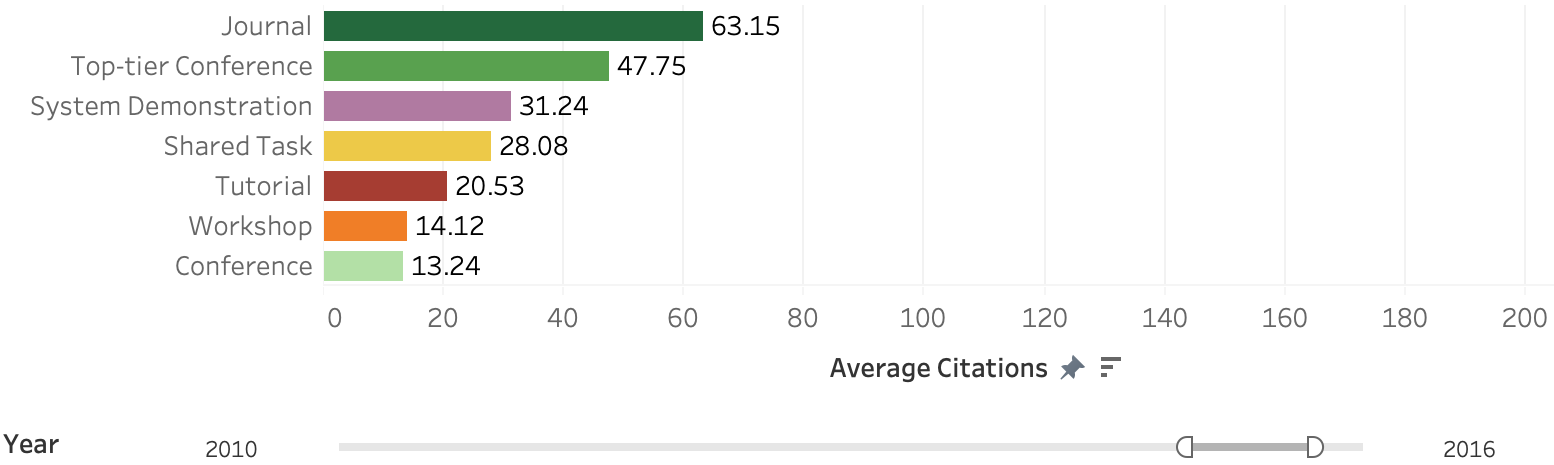}
 	\caption{Average citations by paper type when considering papers published 2010--2016.}
 	\label{fig:PType-AvgCitn-2010-2016}
 \end{center}
 \end{figure*}
 
\subsection{Aggregate Citation Statistics, by Paper Type and Venue}

\noindent \textit{Q. What are the average number of citations received by different types of papers: main conference papers, workshop papers, student research papers, shared task papers, and system demonstration papers?}\\

\begin{figure*}[t!]
 \begin{center}
 	\includegraphics[width=0.9\columnwidth]{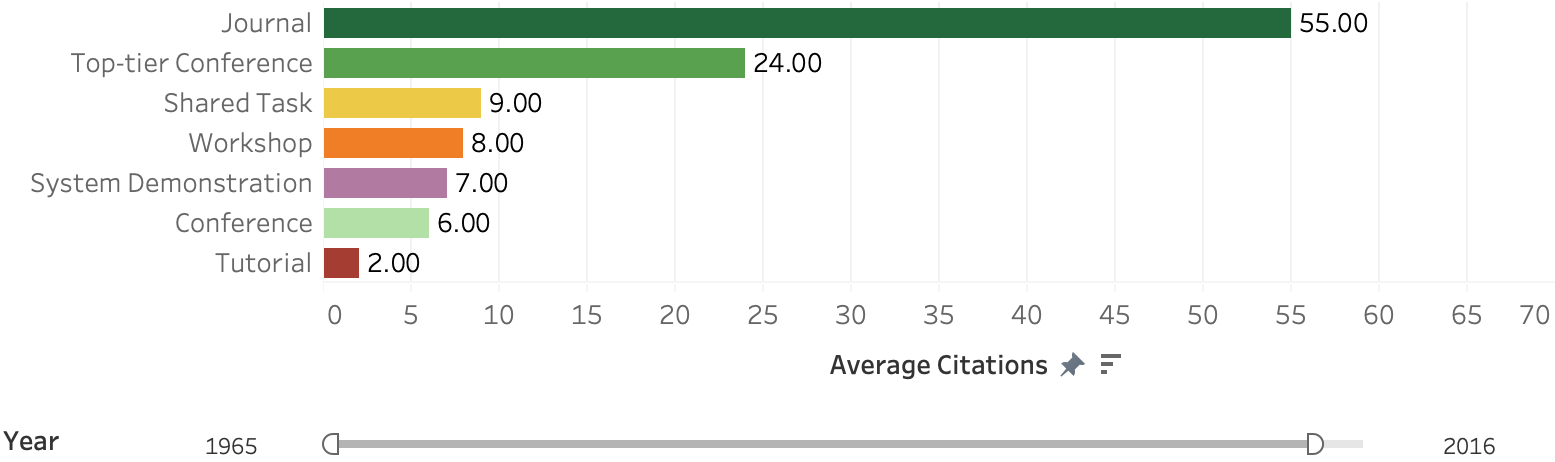}
 	\caption{Median citations by paper type when considering papers published 1965--2016}
 	\label{fig:PType-MedianCitn-1965-2016}
 \end{center}
 \end{figure*}

\begin{figure*}[t!]
 \begin{center}
 	\includegraphics[width=0.9\columnwidth]{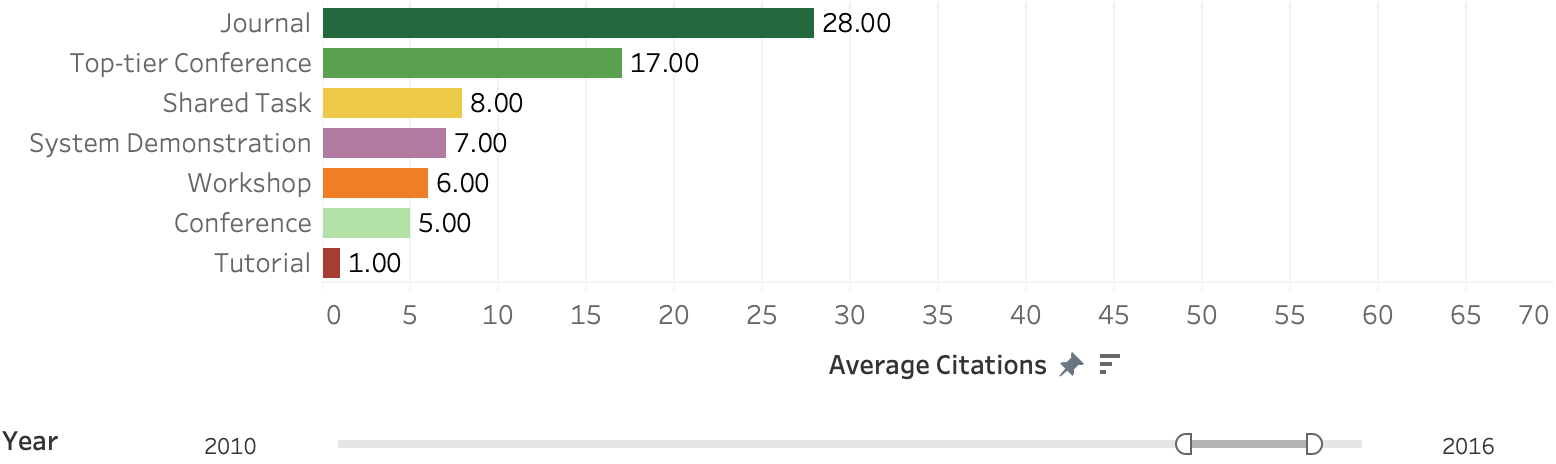}
 	\caption{Median citations by paper type when considering papers published 2010--2016.}
 	\label{fig:PType-MedianCitn-2010-2016}
 \end{center}
 \end{figure*}

\noindent A. In this analysis, we include only those AA’ papers that were published in 2016 or earlier (to allow for at least 2.5 years to collect citations). There are 26,949 such papers.
Figures \ref{fig:PType-AvgCitn-1965-2016} and \ref{fig:PType-AvgCitn-2010-2016} show the average citations by paper type when considering papers published 1965–2016 and 2010–2016, respectively.
Figures \ref{fig:PType-MedianCitn-1965-2016} and \ref{fig:PType-MedianCitn-2010-2016} show the medians.


\noindent Discussion: Journal papers have much higher average and median citations than other papers, but the gap between them and top-tier conferences is markedly reduced when considering papers published since 2010.

System demo papers have the third highest average citations; however, shared task papers have the third highest median citations. The popularity of shared tasks and the general importance given to beating the state of the art (SOTA) seems to have grown in recent years---something that has come under criticism.

It is interesting to note that in terms of citations, workshop papers are doing somewhat better than the conferences that are not top tier.
Finally, the citation numbers for tutorials show that even though a small number of tutorials are well cited, a majority receive 1 or no citations. This is in contrast to system demo papers that have average and median citations that are higher or comparable to workshop papers.

    Throughout the analyses in this article, we see that median citation numbers are markedly lower than average citation numbers. This is particularly telling. It shows that while there are some very highly cited papers, a majority of the papers obtain much lower number of citations---and when considering papers other than journals and top-tier conferences, the number of citations is frequently lower than ten.\\

\noindent \textit{Q. What are the average number of citations received by the long and short ACL main conference papers, respectively?}\\

\begin{figure*}[t!]
 \begin{center}
 \includegraphics[width=0.7\columnwidth]{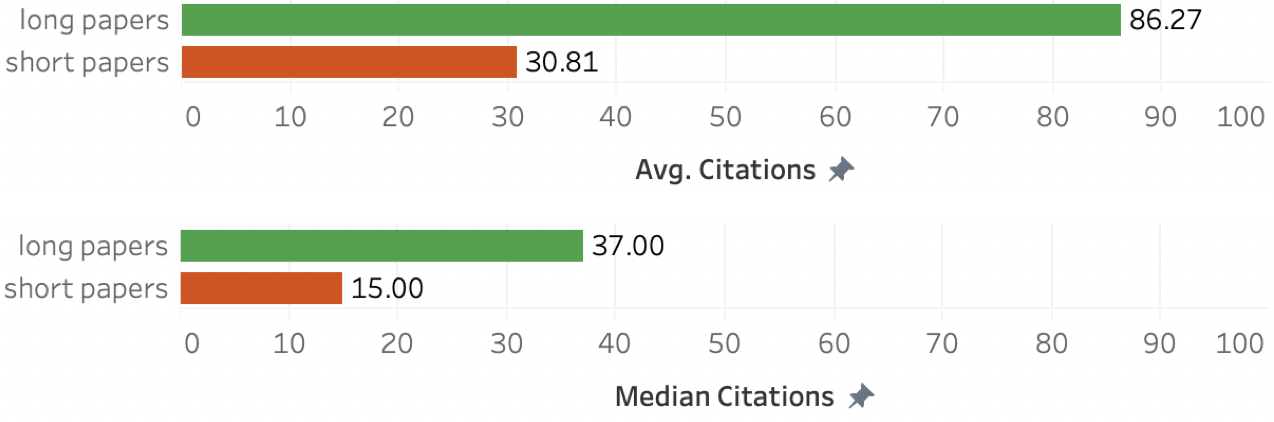}
 	\caption{Average and median citations for long and short papers.}
 	\label{fig:Average-Citn-Long-short}
 \end{center}
 \end{figure*}

\noindent A. Short papers were introduced at ACL in 2003. Since then ACL is by far the venue with the most number of short papers (compared to other venues). So we compare long and short papers published at ACL since 2003 to determine their average citations. Once again, we limit the papers to those published until 2016 to allow for the papers to have time to collect citations.
Figure \ref{fig:Average-Citn-Long-short} shows the average and median citations for long and short papers.\\

\noindent Discussion: On average, long papers get almost three times as many citations as short papers. However, the median for long papers is two-and-half times that of short papers. This difference might be because some very heavily cited long papers push the average up for long papers.\\

\noindent Q. \textit{Which venue has publications with the highest average number of citations? What is the average number of citations for ACL and EMNLP papers? What is this average for other venues? What are the average citations for workshop papers, system demonstration papers, and shared task papers?}\\

\begin{figure*}[t!]
 \begin{center}
 \includegraphics[width=\columnwidth]{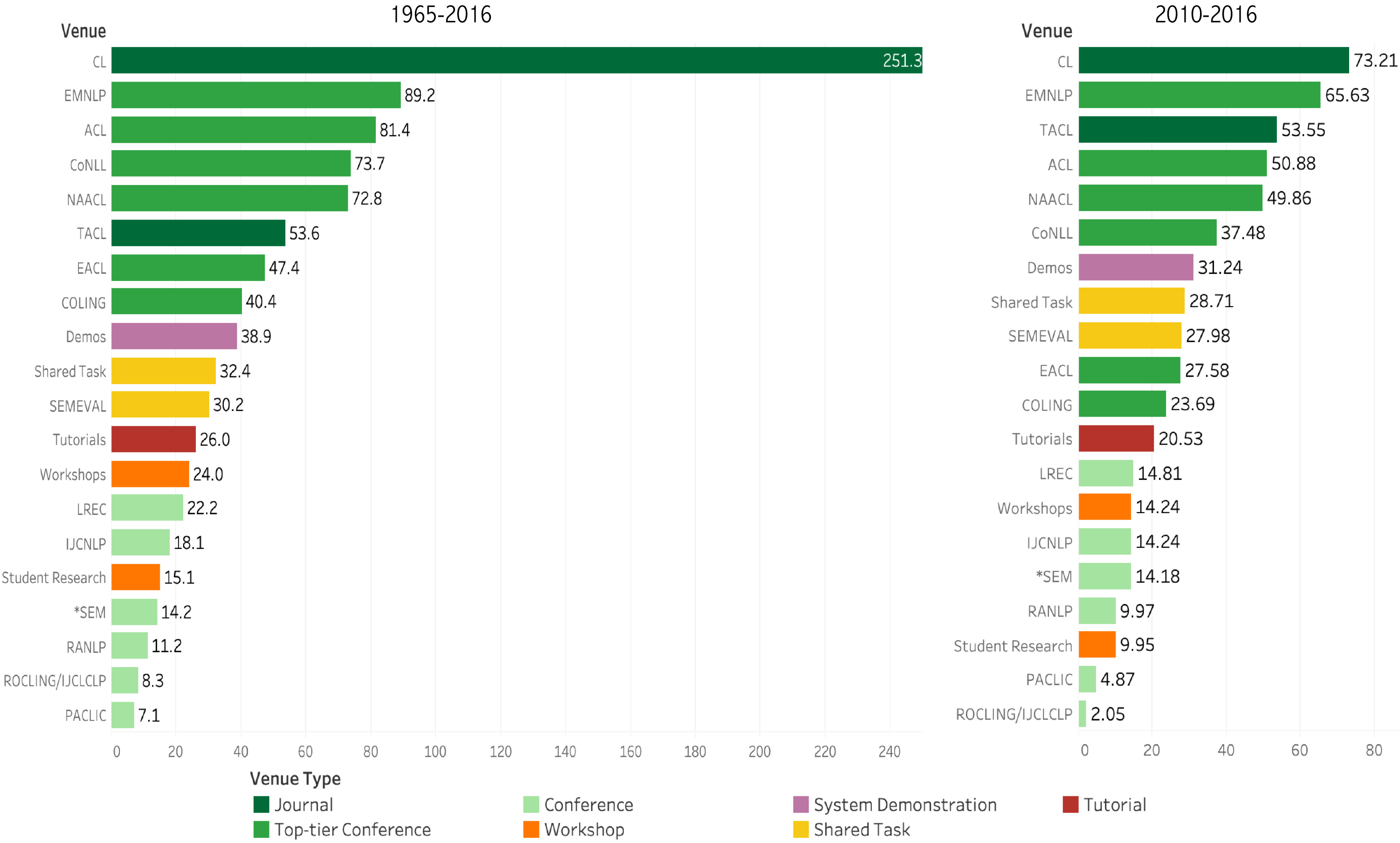}
 	\caption{Average  citations for papers published 1965--2016 (left side) and 2010--2016 (right side), grouped by venue and paper type.}
 	\label{fig:Venue-AvgCitn-1965-2010-2016}
 \end{center}
 \vspace*{-6mm}
 \end{figure*}

\noindent A. CL journal has the highest average citations per paper. Figure \ref{fig:Venue-AvgCitn-1965-2010-2016} 
shows the average citations for AA’ papers published 1965--2016 and 2010--2016, respectively, grouped by venue and paper type. (Figure with median citations is available online.\footnote{https://medium.com/@nlpscholar/the-state-of-nlp-literature-part-iiia-845eb5dc3364}) 


\noindent Discussion: In terms of citations, TACL papers have not been as successful as EMNLP and ACL; however, CL journal (the more traditional journal paper venue) has the highest average and median paper citations (by a large margin). This gap has reduced in papers published since 2010.

When considering papers published between 2010 and 2016, the system demonstration papers, the SemEval shared task papers, and non-SemEval shared task papers have notably high average (surpassing those of EACL and COLING); however their median citations are lower. This is likely because some heavily cited papers have pushed the average up. Nonetheless, it is interesting to note how, in terms of citations, demo and shared task papers have surpassed many conferences and even become competitive with some top-tier conferences such as EACL and COLING.\\

\noindent \textit{Q. What percent of the AA’ papers that were published in 2016 or earlier are cited more than 1000 times? How many more than 10 times? How many papers are cited 0 times?}\\

\begin{figure*}[t!]
 \begin{center}
 	\includegraphics[width=0.35\columnwidth]{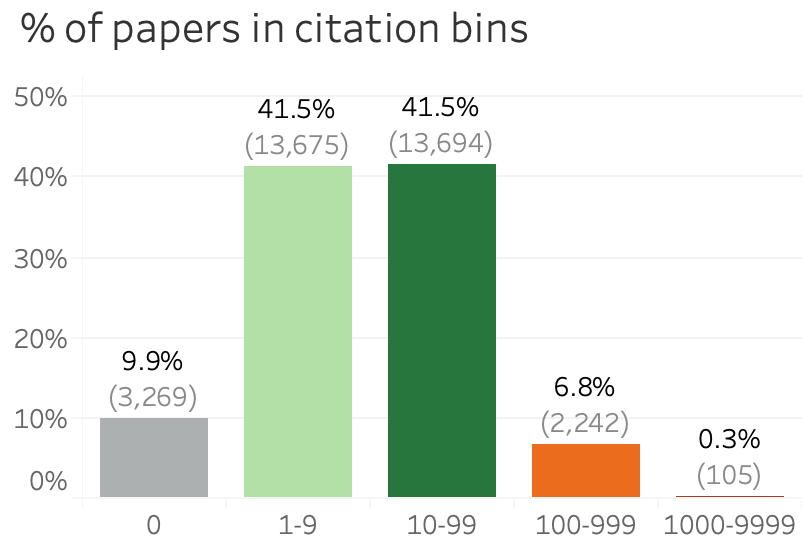}
 	\caption{The percentage of AA’ papers in various citation bins.}
 	\label{fig:citnBins-1965-2016}
 \end{center}
 \vspace*{-6mm}
 \end{figure*}

\noindent A. Google Scholar invented the i-10 index as another measure of author research impact. It stands for the number of papers by an author that received ten or more citations. (Ten here is somewhat arbitrary, but reasonable.) Similar to that, one can look at the impact of AA’ as a whole and the impact of various subsets of AA’ through the number of papers in various citation bins.
Figure \ref{fig:citnBins-1965-2016} shows the percentage of AA’ papers in various citation bins.
(The percentages of papers when considering papers from specific time spans are available online.)

\noindent Discussion: About 56\% of the papers are cited ten or more times. 6.4\% of the papers are never cited. Note also that some portion of the 1–9 bin likely includes papers that only received self-citations.
It is interesting that the percentage of papers with 0 citations is rather steady (between 7.4\% and 8.7\%) for the 1965–1989, 1990–1999, and 2010–2016 periods. The majority of the papers lie in the 10 to 99 citations bin, for all except the recent periods (2010–2016 and 2016Jan–2016Dec). With time, the recent period should also have the majority of the papers in the 10 to 99 citations bin.

The numbers for the 2016Jan–2016Dec papers show that after 2.5 years, about 89\% of the papers have at least one citation and about 33\% of the papers have ten or more citations.\\

\noindent \textit{Q. What are the citation bin percentages for individual venues and paper types?}\\

\begin{figure*}[t!]
 \begin{center}
 	\includegraphics[width=\columnwidth]{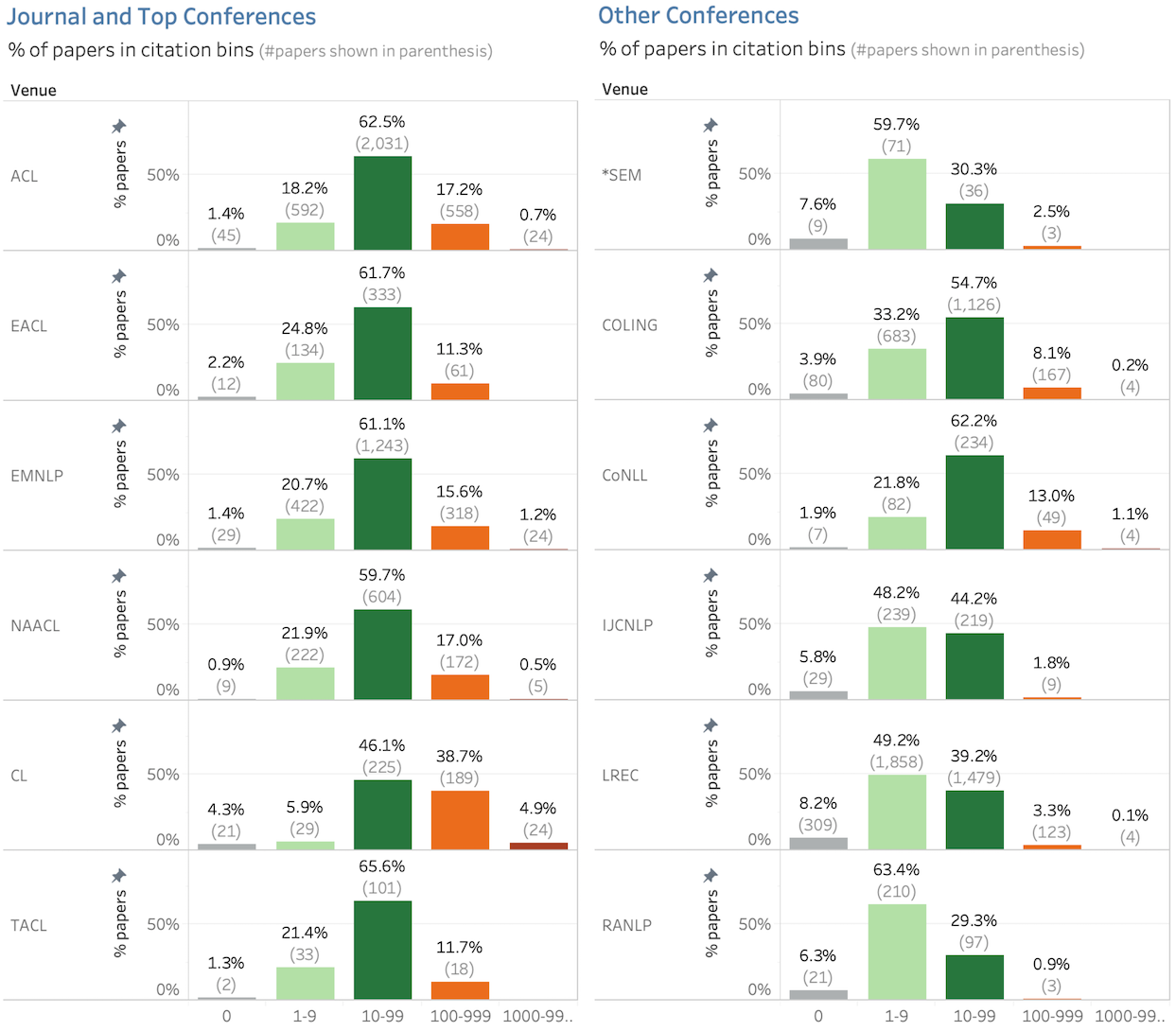}
 	\caption{The citation bin percentages for individual venues and paper types.}
 	\label{fig:21}
 \end{center}
 \end{figure*}

\noindent A. See Figure \ref{fig:21}.\\

\noindent Discussion: Observe that 70 to 80\% of the papers in journals and top-tier conferences have ten or more citations. The percentages are markedly lower (between 30 and 70\%) for the other conferences shown above, and even lower for some other conferences (not shown above).

CL Journal is particularly notable for the largest percentage of papers with 100 or more citations. The somewhat high percentage of papers that are never cited (4.3\%) are likely because some of the book reviews from earlier years are not explicitly marked in CL journal, and thus they were not removed from analysis. Also, letters to editors, which are more common in CL journal, tend to often obtain 0 citations.

CL, EMNLP, and ACL have the best track record for accepting papers that have gone on to receive 1000 or more citations. *Sem, the semantics conference, seems to have notably lower percentage of high-citation papers, even though it has fairly competitive acceptance rates.

Instead of percentage, if one considers raw numbers of papers that have at least ten citations (i-10 index), then LREC is particularly notable in terms of the large number of papers it accepts that have gone on to obtain ten or more citations ($\sim$1600). Thus, by producing a large number of moderate-to-high citation papers, and introducing many first-time authors, LREC is one of the notable (yet perhaps undervalued) engines of impact on NLP.

About 50\% of the SemEval shared task papers received 10 or more citations, and about 46\% of the non-SemEval Shared Task Papers received 10 or more citations. About 47\% of the workshop papers received ten or more citations. About 43\% of the demo papers received 10 or more citations.

\subsection{Citations to Papers by Areas of Research}

\noindent \textit{Q. What is the average number of citations of AA' papers that have machine translation in the title? What about papers that have the term sentiment analysis or word representations?}\\

\begin{figure*}[t!]
 \begin{center}
 	\includegraphics[width=\columnwidth]{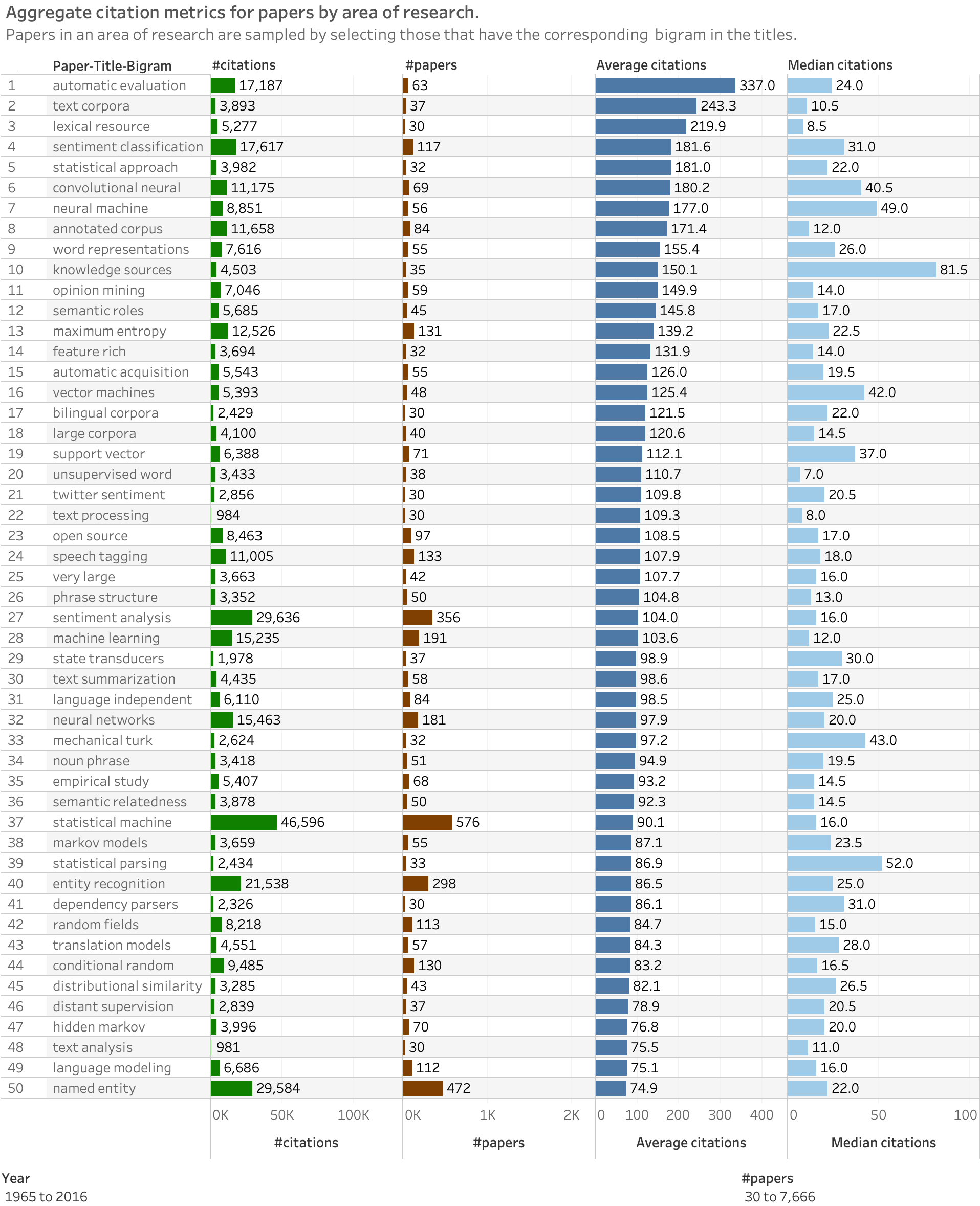}
 	\caption{The top 50 title bigrams ordered by decreasing number of total citations.}
 	\label{fig:bigram-citns}
 \end{center}
 \vspace*{-9mm}
 \end{figure*}

\noindent A. Different areas of research within NLP enjoy varying amounts of attention. In Part II, we looked at the relative popularity of various areas over time---estimated through the number of paper titles that had corresponding terms. (You may also want to see the discussion on the use of paper title terms to sample papers from various, possibly overlapping, areas.) Figure \ref{fig:bigram-citns} shows the top 50 title bigrams ordered by decreasing number of total citations. Only those bigrams that occur in at least 30 AA' papers (published between 1965 and 2016) are considered. (The papers from 2017 and later are not included, to allow for at least 2.5 years for the papers to accumulate citations.)

\noindent  Discussion: The graph shows that the bigram machine translation occurred in 1,659 papers that together accrued more than 93k citations. These papers have on average 68.8 citations and the median citations is 14. Not all machine translation (MT) papers have machine translation in the title. However, arguably, this set of 1,659 papers is a representative enough sample of machine translation papers; and thus, the average and median are estimates of MT in general. Second in the list are papers with statistical machine in the title---most commonly from the phrase statistical machine translation. One expects considerable overlap in the papers across the sets of papers with machine translation and statistical machine, but machine translation likely covers a broader range of research including work before statistical MT was introduced, neural MT, and MT evaluation.

There are fewer papers with sentiment analysis in the title (356), but these have acquired citations at a higher average (104) than both machine translation and statistical machine. The bigram automatic evaluation jumps out because of its high average citations (337). Some of the neural-related bigrams have high median citations, for example, neural machine (49) and convolutional neural (40.5).\\

\begin{figure*}[t!]
 \begin{center}
 	\includegraphics[width=\columnwidth]{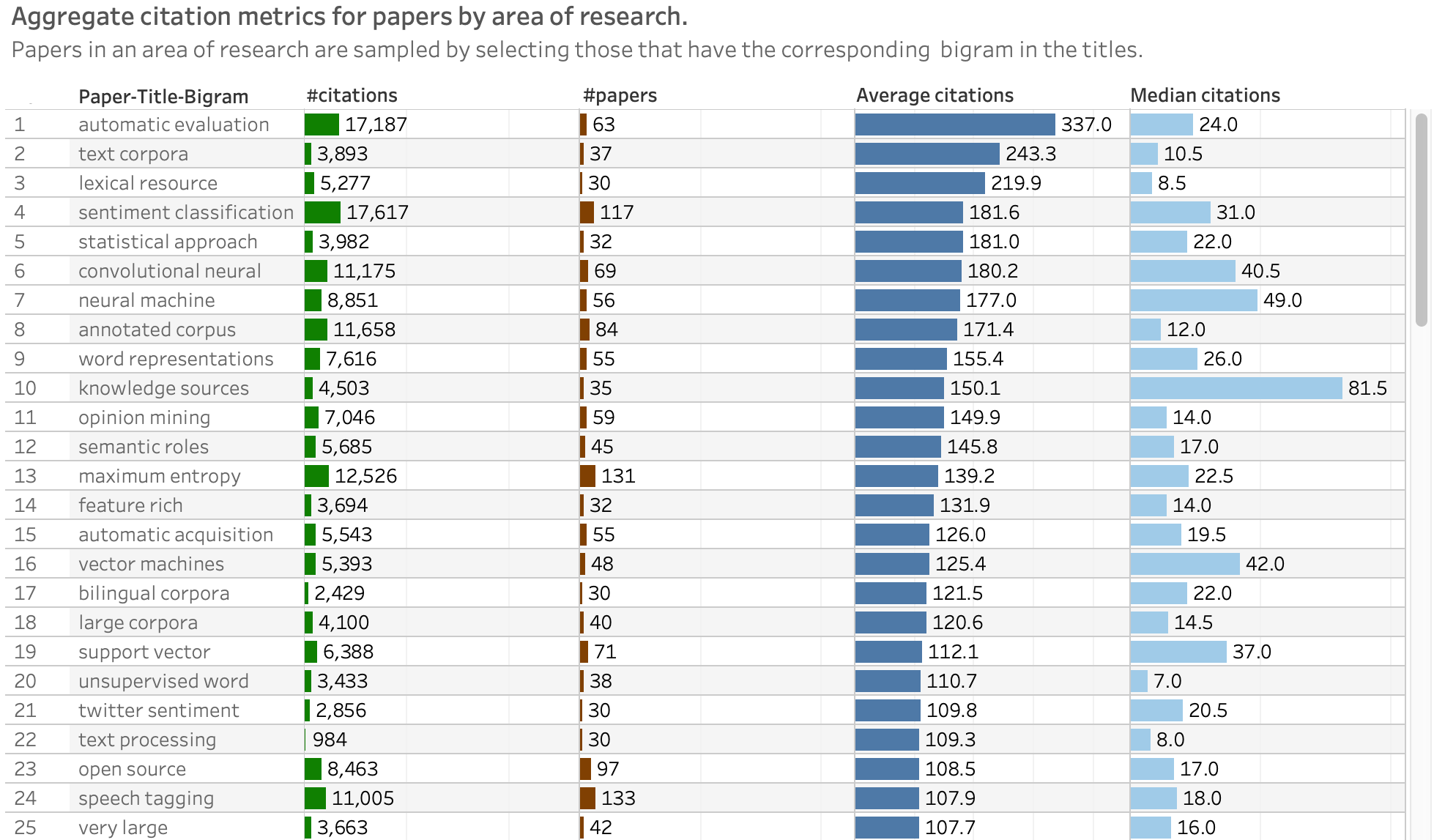}
 	\caption{The lists of top 25 bigrams ordered by average citations.}
 	\label{fig:bigram-citns-by-avg}
 \end{center}
 \vspace*{-3mm}
 \end{figure*}

\noindent Figure \ref{fig:bigram-citns-by-avg} shows the lists of top 25 bigrams ordered by average citations.

\noindent Discussion: Observe the wide variety of topics covered by this list. In some ways that is reassuring for the health of the field as a whole; however, this list does not show which areas are not receiving sufficient attention. It is less clear to me how to highlight those, as simply showing the bottom 50 bigrams by average citations is not meaningful. Also note that this is not in any way an endorsement to write papers with these high-citation bigrams in the title. Doing so is of course no guarantee of receiving a large number of citations.


    



\section{Correlation of Age and Gender with Citations}

In this section, we examine citations across two demographic dimensions: Academic age (number of years one has been publishing) and
    Gender. There are good reasons to study citations across each of these dimensions including, but not limited to, the following:
\begin{itemize}
    \item Areas of research: To better understand research contributions in the context of the area where the contribution is made.
    \item Academic age: To better understand how the challenges faced by researchers at various stages of their career may impact the citations of their papers. For example, how well-cited are first-time NLP authors? On average, at what academic age do citations peak? etc.
    \item Gender: To better understand the extent to which systematic biases (explicit and implicit) pervasive in society and scientific publishing impact author citations.
\end{itemize}

Some of these aspects of study may seem controversial. So it is worth addressing that first. The goal here is not to perpetuate stereotypes about age, gender, or even areas of research. The history of scientific discovery is awash with plenty of examples of bad science that has tried to erroneously show that one group of people is ``better'' than another, with devastating consequences.

    People are far more alike than different. However, different demographic groups have faced (and continue to face) various socio-cultural inequities and biases. Gender and race studies look at how demographic differences shape our experiences. They examine the roles of social institutions in maintaining the inequities and biases.

This work is in support of those studies. Unless we measure differences in outcomes such as scientific productivity and impact across demographic groups, we will not fully know the extent to which these inequities and biases impact our scientific community; and we cannot track the effectiveness of measures to make our universities, research labs, and conferences more inclusive, equitable, and fair.

\subsection{Correlation of Academic Age with Citations}

We introduced NLP academic age earlier in the paper, where
we defined NLP academic age as the number of years one has been publishing in AA.
Here we examine whether NLP academic age impacts citations.
The analyses are done in terms of the academic age of the first author; however, similar analyses can be done for the last author and all authors. (There are limitations to each of these analyses though as discussed further below.)

First author is a privileged position in the author list as it is usually reserved for the researcher that has done the most work and writing. The first author is also usually the main driver of the project; although, their mentor or advisor may also be a significant driver of the project. Sometimes multiple authors may be marked as first authors in the paper, but the current analysis simply takes the first author from the author list. In many academic communities, the last author position is reserved for the most senior or mentoring researcher. However, in non-university research labs and in large collaboration projects, the meaning of the last author position is less clear. (Personally, I prefer author names ordered by the amount of work done.)

Examining all authors is slightly more tricky as one has to decide how to credit the citations to the possibly multiple authors. It might also not be a clear indicator of differences across gender as a large number of the papers in AA have both male and female authors.\\

\noindent \textit{Q. How does the NLP academic age of the first author correlate with the amount of citations? Are first-year authors less cited than those with more experience?}\\

\begin{figure*}[t!]
 \begin{center}
 	\includegraphics[width=0.88\columnwidth]{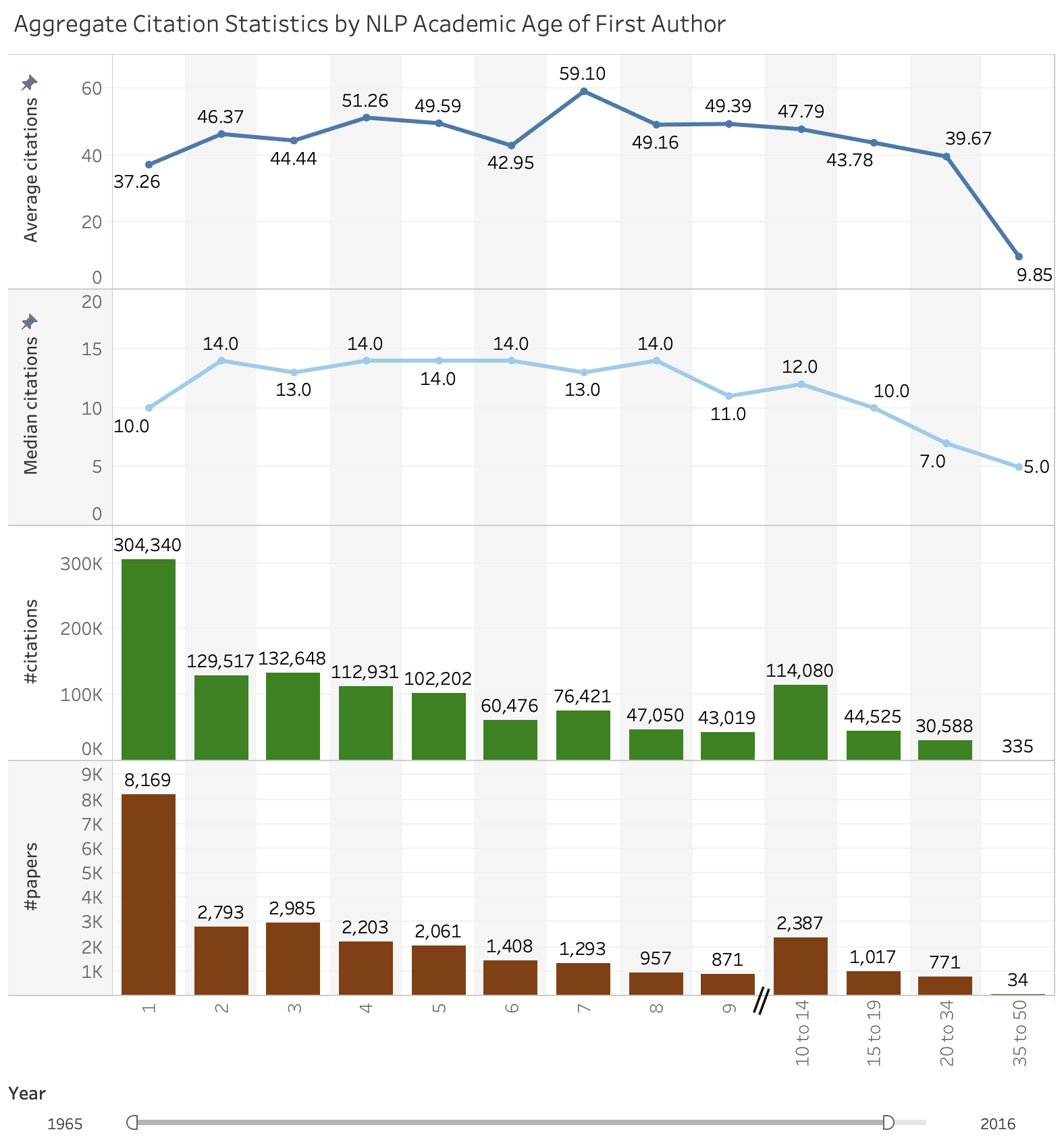}
 	\caption{Aggregate citation statistics by academic age.}
 	\label{fig:Citns by Age}
 \end{center}
 \end{figure*}

\noindent A. Figure \ref{fig:Citns by Age} shows various aggregate citation statistics corresponding to academic age. To produce the graph we put each paper in a bin corresponding to the academic age of the first author when the paper was published. For example, if the first author of a paper had an academic age of 3 when that paper was published, then the paper goes in bin 3. We then calculate \#papers, \#citations, median citations, and average citations for each bin. 
For the figure below, We further group the bins 10 to 14, 15 to 19, 20 to 34, and 35 to 50. These groupings are done to avoid clutter, and also because many of the higher age bins have a low number of papers.

\noindent Discussion: Observe that the number of papers where the first author has academic age 1 is much larger than the number of papers in any other bin. This is largely because a large number of authors in AA have written exactly one paper as first author. Also, about 60\% of the authors in AA (17,874 out of the 29,941 authors) have written exactly one paper (regardless of author position).

The curves for the average and median citations have a slight upside down U shape. The relatively lower average and median citations in year 1 (37.26 and 10, respectively) indicate that being new to the field has some negative impact on citations. The average increases steadily from year 1 to year 4, but the median is already at the highest point by year 2. One might say, that year 2 to year 14 are the period of steady and high citations. Year 15 onwards, there is a steady decline in the citations. It is probably wise to not draw too many conclusions from the averages of the 35 to 50 bin, because of the small number of papers. There seems to be a peak in average citations at age 7. However, there is not a corresponding peak in the median. Thus the peak in average might be due to an increase in the number of very highly cited papers.
Citations to Papers by First Author Gender

As noted in Part I, neither ACL nor the ACL Anthology have recorded demographic information for the vast majority of the authors. Thus we use the same setup discussed earlier in the section on demographics, to determine gender using the United States Social Security Administration database of names and genders of newborns to identify 55,133 first names that are strongly associated with females (probability $\geq$99\%) and 29,873 first names that are strongly associated with males (probability $\geq$99\%). \\


\noindent \textit{Q. On average, are women cited less than men?}\\

\begin{figure*}[t!]
 \begin{center}
 	\includegraphics[width=\columnwidth]{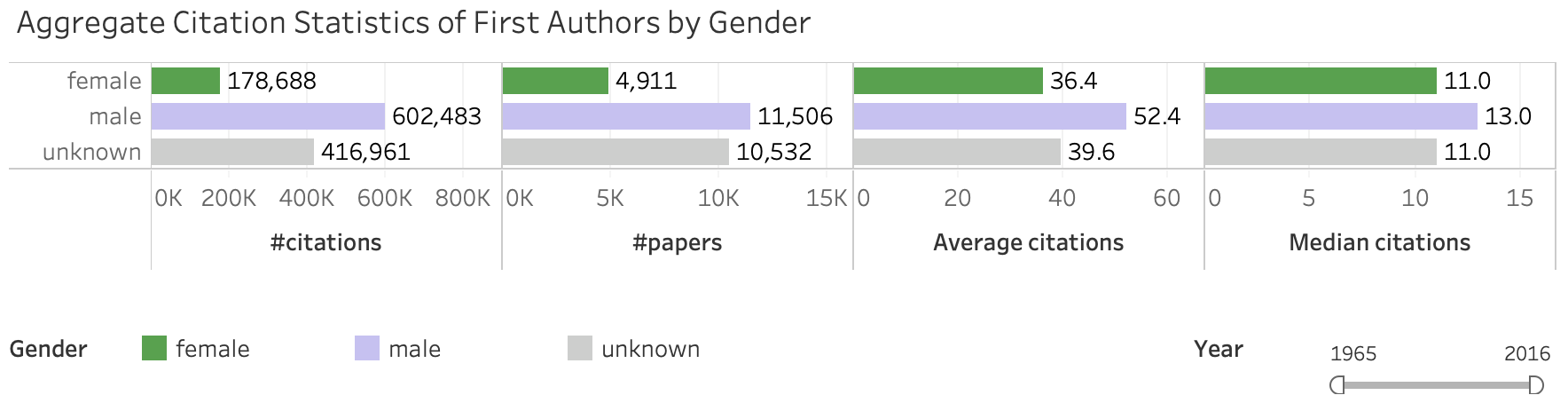}
 	\caption{Average citations received by female and male first authors.}
 	\label{fig:citns-gender}
 \end{center}
 \end{figure*}

\noindent A. Yes, on average, female first author papers have received markedly fewer citations than male first author papers (36.4 compared to 52.4). The difference in median is smaller (11 compared to 13). See Figure \ref{fig:citns-gender}.

\noindent Discussion: The large difference in averages and smaller difference in medians suggests that there are markedly more very heavily cited male first-author papers than female first-author papers.
The gender-unknown category, which here largely consist of authors with Chinese origin names and names that are less strongly associated with one gender have a slightly higher average, but the same median citations, as authors with female-associated first names.

\noindent The differences in citations, or \textit{citation gap}, across genders may: (1) vary by period of time; (2) vary due to confounding factors such as academic age and areas of research. We explore these next.\\

\noindent \textit{Q. How has the citation gap across genders changed over the years?}\\

\begin{figure*}[t!]
 \begin{center}
 \includegraphics[width=\columnwidth]{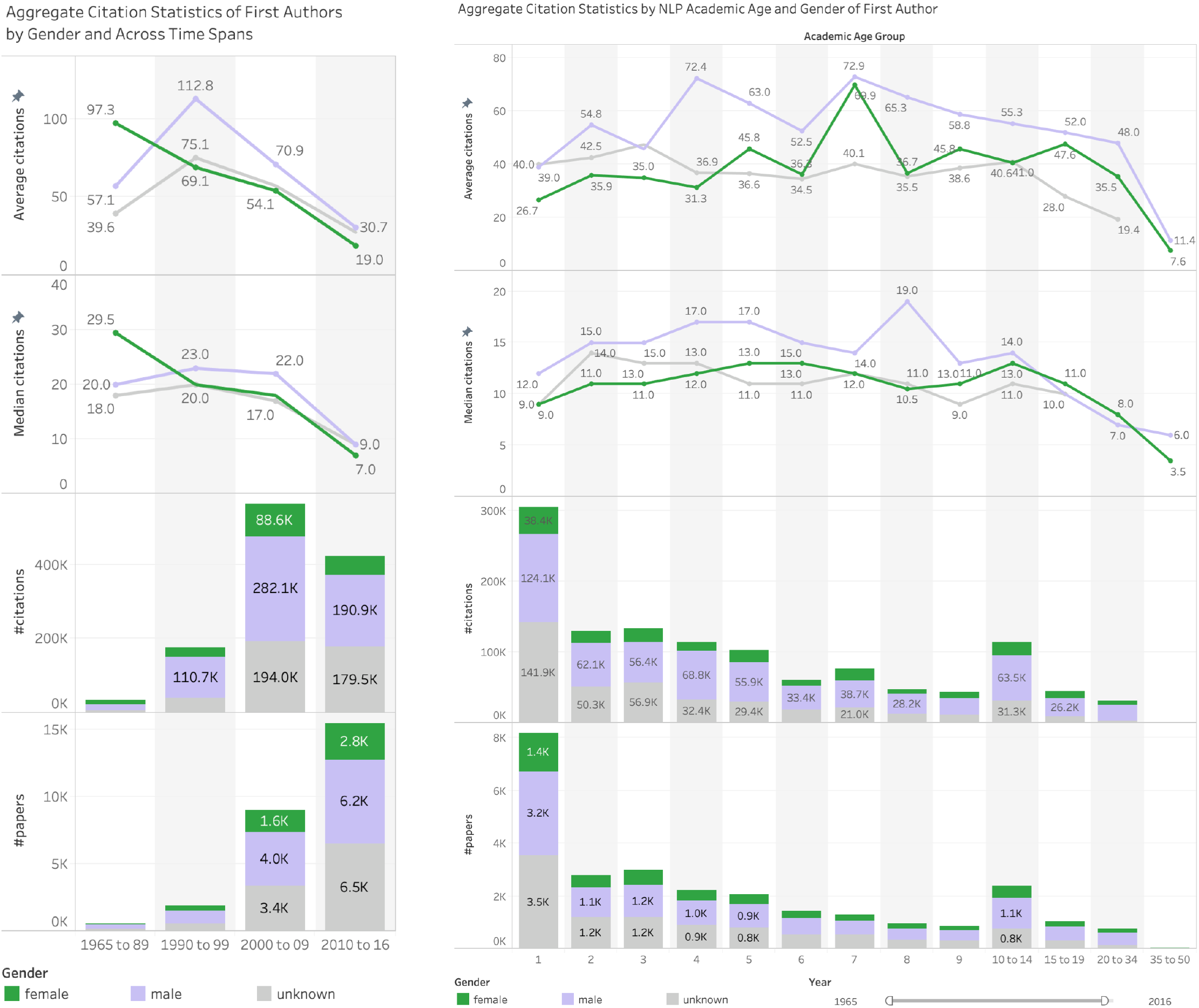}
 	\caption{Citation gap across genders for papers: (a) published in different time spans, (b) by academic age.}
 	\label{fig:citn-gender-time-band}
 \end{center}
 \vspace*{-6mm}
 \end{figure*}

\noindent A. Figure \ref{fig:citn-gender-time-band} (left side) shows the citation statistics across four time periods.\\

\noindent Discussion: Observe that female first authors have always been a minority in the history of ACL; however, on average, their papers from the early years (1965 to 1989) received a markedly higher number of citations than those of male first authors from the same period. We can see from the graph that this changed in the 1990s where male first-author papers obtained markedly more citations on average. The citation gap reduced considerably in the 2000s, and the 2010–2016 period saw a further slight reduction in the citation gap.

It is also interesting to note that the gender-unknown category has almost bridged the gap with the males in this most recent time period. Further, the proportion of the gender-unknown authors has increased over the years---arguably, an indication of better representations of authors from around the world in recent years. (Nonetheless, as indicated in Part I, there is still plenty to be done to promote greater inclusion of authors from Africa and South America.)\\

\noindent \textit{Q. How have citations varied by gender and academic age? Are women less cited  because of a greater proportion of new-to-NLP female first authors than new-to-NLP male first authors?}\\

\noindent A. Figure \ref{fig:citn-gender-time-band} (right side) shows citation statistics broken down by gender and academic age. (This figure is similar to the academic age graph seen earlier, except that it shows separate average and median lines for female, male, and unknown gender first authors.)

\noindent Discussion: The graphs show that female first authors consistently receive fewer citations than male authors for the first fifteen years. The trend is inverted with a small citation gap in the 15th to 34th years period.\\


\noindent \textit{Q. Is the citation gap common across the vast majority of areas of research within NLP? Is the gap simply because more women work in areas that receive low numbers of citations (regardless of gender)?}\\

\begin{figure*}[t]
 \begin{center}
 	\includegraphics[width=\columnwidth]{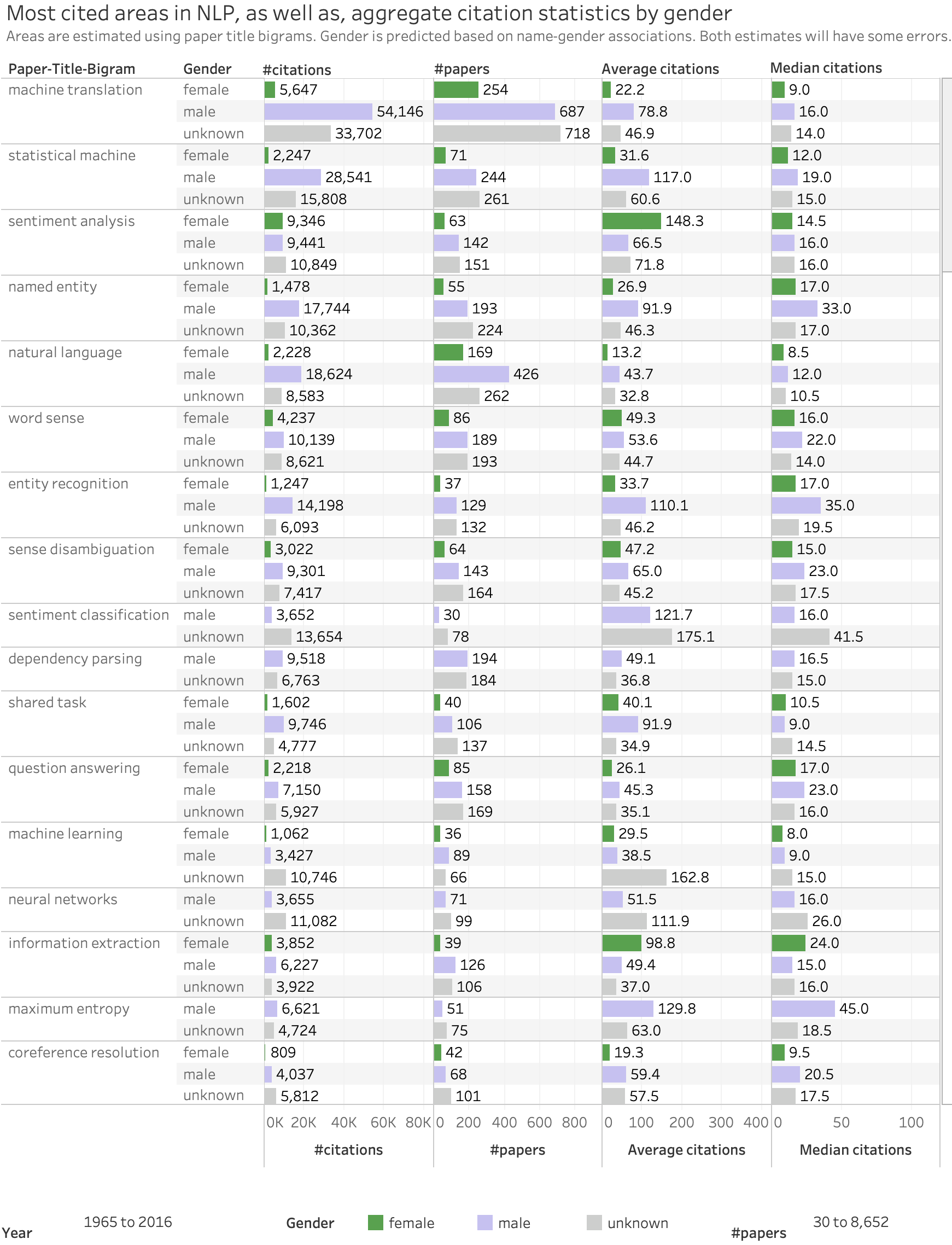}
 	\caption{The most cited areas of research along with citation statistics split by gender of the first authors of corresponding papers.}
 	\label{fig:citn-areas-gender}
 \end{center}
 \end{figure*}

\noindent A. Figure \ref{fig:citn-areas-gender} shows the most cited areas of research along with citation statistics split by gender of the first authors of corresponding papers. (This figure is similar to the areas of research graph seen earlier, except that it shows separate citation statistics for the genders.) Note that the figure includes rows for only those bigram and gender pairs with at least 30 AA’ papers (published between 1965 and 2016). Thus for some of the bigrams certain gender entries are not shown.

\noindent Discussion: Numbers for an additional 32 areas are available online.\footnote{https://medium.com/@nlpscholar/about-nlp-scholar-62cb3b0f4488} 
Observe that in only about 12\% (7 of the top 59) of the most cited areas of research, women received higher average citations than men. These include: sentiment analysis, information extraction, document summarization, spoken dialogue, cross lingual (research), dialogue, systems, language generation. (Of course, note that some of the 59 areas, as estimated using title term bigrams, are overlapping. Also, we did not include \textit{large scale} in the list above because the difference in averages is very small and it is not really an area of research.)
Thus, the citation gap is common across a majority of the high-citations areas within NLP.

\section{Conclusions}
This work examined the ACL Anthology to identify broad trends in productivity, focus, and impact. We examined several questions such as: who and how many of us are publishing? what are we publishing on? where and in what form are we publishing? and what is the impact of our publications? Particular attention was paid to the demographics and inclusiveness of the NLP community. Notably, we showed that only about 30\% of first authors are female, and that this percentage has not improved since the year 2000. We also showed that, on average, female first authors are cited less than male first authors, even when controlling for academic age. We hope that recording citation and participation gaps across demographic groups will encourage our university, industry, and government research labs to be more inclusive and fair. 
Several additional aspects of the AA will be explored in future work (see the bottom of the blog posts).\footnote{https://medium.com/@nlpscholar/state-of-nlp-cbf768492f90}\\\\

\noindent \textbf{Acknowledgments}\\
This work was possible due to the helpful discussion and encouragement from a number of awesome people, including: Dan Jurafsky, Tara Small, Michael Strube, Cyril Goutte, Eric Joanis, Matt Post, Patrick Littell, Torsten Zesch, Ellen Riloff, Norm Vinson, Iryna Gurevych, Rebecca Knowles, Isar Nejadgholi, and Peter Turney. Also, a big thanks to the ACL Anthology team for creating and maintaining a wonderful resource. 

\newpage

\starttwocolumn
\bibliography{state-of-NLP}

\begin{thebibliography}{22}
\expandafter\ifx\csname natexlab\endcsname\relax\def\natexlab#1{#1}\fi

\bibitem[{Anderson, McFarland, and Jurafsky(2012)}]{anderson2012towards}
Anderson, Ashton, Dan McFarland, and Dan Jurafsky. 2012.
\newblock Towards a computational history of the acl: 1980-2008.
\newblock In \emph{Proceedings of the ACL-2012 Special Workshop on
  Rediscovering 50 Years of Discoveries}, pages 13--21, Association for
  Computational Linguistics.

\bibitem[{Aya, Lagoze, and Joachims(2005)}]{aya2005citation}
Aya, Selcuk, Carl Lagoze, and Thorsten Joachims. 2005.
\newblock Citation classification and its applications.
\newblock In \emph{Knowledge Management: Nurturing Culture, Innovation, and
  Technology}. World Scientific, pages 287--298.

\bibitem[{Bird et~al.(2008)Bird, Dale, Dorr, Gibson, Joseph, Kan, Lee, Powley,
  Radev, and Tan}]{bird2008acl}
Bird, Steven, Robert Dale, Bonnie~J Dorr, Bryan Gibson, Mark~Thomas Joseph,
  Min-Yen Kan, Dongwon Lee, Brett Powley, Dragomir~R Radev, and Yee~Fan Tan.
  2008.
\newblock The acl anthology reference corpus: A reference dataset for
  bibliographic research in computational linguistics.

\bibitem[{Bos and Nitza(2019)}]{bos2019interdisciplinary}
Bos, Arthur~R and Sandrine Nitza. 2019.
\newblock Interdisciplinary comparison of scientific impact of publications
  using the citation-ratio.
\newblock \emph{Data Science Journal}, 18(1).

\bibitem[{Bulaitis(2017)}]{bulaitis2017measuring}
Bulaitis, Zoe. 2017.
\newblock Measuring impact in the humanities: Learning from accountability and
  economics in a contemporary history of cultural value.
\newblock \emph{Palgrave Communications}, 3(1):7.

\bibitem[{Howland(2010)}]{howland2010scholarly}
Howland, Jared~L. 2010.
\newblock How scholarly is google scholar? a comparison to library databases.

\bibitem[{Ioannidis et~al.(2019)Ioannidis, Baas, Klavans, and
  Boyack}]{ioannidis2019standardized}
Ioannidis, John~PA, Jeroen Baas, Richard Klavans, and Kevin~W Boyack. 2019.
\newblock A standardized citation metrics author database annotated for
  scientific field.
\newblock \emph{PLoS biology}, 17(8):e3000384.

\bibitem[{Khabsa and Giles(2014)}]{khabsa2014number}
Khabsa, Madian and C~Lee Giles. 2014.
\newblock The number of scholarly documents on the public web.
\newblock \emph{PloS one}, 9(5):e93949.

\bibitem[{Mariani, Francopoulo, and Paroubek(2018)}]{mariani2018nlp4nlp}
Mariani, Joseph, Gil Francopoulo, and Patrick Paroubek. 2018.
\newblock The nlp4nlp corpus (i): 50 years of publication, collaboration and
  citation in speech and language processing.
\newblock \emph{Frontiers in Research Metrics and Analytics}, 3:36.

\bibitem[{Mart{\'\i}n-Mart{\'\i}n et~al.(2018)Mart{\'\i}n-Mart{\'\i}n,
  Orduna-Malea, Thelwall, and L{\'o}pez-C{\'o}zar}]{martin2018google}
Mart{\'\i}n-Mart{\'\i}n, Alberto, Enrique Orduna-Malea, Mike Thelwall, and
  Emilio~Delgado L{\'o}pez-C{\'o}zar. 2018.
\newblock Google scholar, web of science, and scopus: A systematic comparison
  of citations in 252 subject categories.
\newblock \emph{Journal of Informetrics}, 12(4):1160--1177.

\bibitem[{Mohammad et~al.(2009)Mohammad, Dorr, Egan, Hassan, Muthukrishan,
  Qazvinian, Radev, and Zajic}]{mohammad2009using}
Mohammad, Saif, Bonnie Dorr, Melissa Egan, Ahmed Hassan, Pradeep Muthukrishan,
  Vahed Qazvinian, Dragomir Radev, and David Zajic. 2009.
\newblock Using citations to generate surveys of scientific paradigms.
\newblock In \emph{Proceedings of human language technologies: The 2009 annual
  conference of the North American chapter of the association for computational
  linguistics}, pages 584--592, Association for Computational Linguistics.

\bibitem[{Nanba, Kando, and Okumura(2011)}]{nanba2011classification}
Nanba, Hidetsugu, Noriko Kando, and Manabu Okumura. 2011.
\newblock Classification of research papers using citation links and citation
  types: Towards automatic review article generation.
\newblock \emph{Advances in Classification Research Online}, 11(1):117--134.

\bibitem[{Ordu{\~n}a-Malea et~al.(2014)Ordu{\~n}a-Malea, Ayll{\'o}n,
  Mart{\'\i}n-Mart{\'\i}n, and L{\'o}pez-C{\'o}zar}]{orduna2014size}
Ordu{\~n}a-Malea, Enrique, Juan~Manuel Ayll{\'o}n, Alberto
  Mart{\'\i}n-Mart{\'\i}n, and Emilio~Delgado L{\'o}pez-C{\'o}zar. 2014.
\newblock About the size of google scholar: playing the numbers.
\newblock \emph{arXiv preprint arXiv:1407.6239}.

\bibitem[{Pham and Hoffmann(2003)}]{pham2003new}
Pham, Son~Bao and Achim Hoffmann. 2003.
\newblock A new approach for scientific citation classification using cue
  phrases.
\newblock In \emph{Australasian Joint Conference on Artificial Intelligence},
  pages 759--771, Springer.

\bibitem[{Priem and Hemminger(2010)}]{priem2010scientometrics}
Priem, Jason and Bradely~H Hemminger. 2010.
\newblock Scientometrics 2.0: New metrics of scholarly impact on the social
  web.
\newblock \emph{First monday}, 15(7).

\bibitem[{Qazvinian et~al.(2013)Qazvinian, Radev, Mohammad, Dorr, Zajic,
  Whidby, and Moon}]{qazvinian2013generating}
Qazvinian, Vahed, Dragomir~R Radev, Saif~M Mohammad, Bonnie Dorr, David Zajic,
  Michael Whidby, and Taesun Moon. 2013.
\newblock Generating extractive summaries of scientific paradigms.
\newblock \emph{Journal of Artificial Intelligence Research}, 46:165--201.

\bibitem[{Radev et~al.(2016)Radev, Joseph, Gibson, and
  Muthukrishnan}]{radev2016bibliometric}
Radev, Dragomir~R, Mark~Thomas Joseph, Bryan Gibson, and Pradeep Muthukrishnan.
  2016.
\newblock A bibliometric and network analysis of the field of computational
  linguistics.
\newblock \emph{Journal of the Association for Information Science and
  Technology}, 67(3):683--706.

\bibitem[{Ravenscroft et~al.(2017)Ravenscroft, Liakata, Clare, and
  Duma}]{ravenscroft2017measuring}
Ravenscroft, James, Maria Liakata, Amanda Clare, and Daniel Duma. 2017.
\newblock Measuring scientific impact beyond academia: An assessment of
  existing impact metrics and proposed improvements.
\newblock \emph{PloS one}, 12(3):e0173152.

\bibitem[{Schluter(2018)}]{schluter2018glass}
Schluter, Natalie. 2018.
\newblock The glass ceiling in nlp.
\newblock In \emph{Proceedings of the 2018 Conference on Empirical Methods in
  Natural Language Processing}, pages 2793--2798.

\bibitem[{Teufel, Siddharthan, and Tidhar(2006)}]{teufel2006automatic}
Teufel, Simone, Advaith Siddharthan, and Dan Tidhar. 2006.
\newblock Automatic classification of citation function.
\newblock In \emph{Proceedings of the 2006 conference on empirical methods in
  natural language processing}, pages 103--110, Association for Computational
  Linguistics.

\bibitem[{Yogatama et~al.(2011)Yogatama, Heilman, O'Connor, Dyer, Routledge,
  and Smith}]{yogatama2011predicting}
Yogatama, Dani, Michael Heilman, Brendan O'Connor, Chris Dyer, Bryan~R
  Routledge, and Noah~A Smith. 2011.
\newblock Predicting a scientific community's response to an article.
\newblock In \emph{Proceedings of the Conference on Empirical Methods in
  Natural Language Processing}, pages 594--604, Association for Computational
  Linguistics.

\bibitem[{Zhu et~al.(2015)Zhu, Turney, Lemire, and Vellino}]{zhu2015measuring}
Zhu, Xiaodan, Peter Turney, Daniel Lemire, and Andr{\'e} Vellino. 2015.
\newblock Measuring academic influence: Not all citations are equal.
\newblock \emph{Journal of the Association for Information Science and
  Technology}, 66(2):408--427.

\end{thebibliography}

\end{document}